\documentclass[acmsmall]{acmart}

\usepackage{booktabs} 
\usepackage{subcaption}
\usepackage{mathtools}
\usepackage{multicol}
\usepackage{color,soul}
\sethlcolor{white}
\usepackage{makecell}
\usepackage{colortbl}
\usepackage{tabularx,colortbl}
\usepackage{caption}
\usepackage{booktabs}
\usepackage{multirow}
\usepackage{tikz}
\usepackage{multicol}
\usepackage{enumitem}
\usepackage{url}
\usepackage{graphics}
\usepackage{float}

\begin{document}

\setcopyright{acmcopyright}
\acmJournal{PACMHCI}
\acmYear{2022} \acmVolume{6} \acmNumber{CSCW1} \acmArticle{66} \acmMonth{4} \acmPrice{15.00}\acmDOI{10.1145/3512913}

\title{The Effects of System Initiative during Conversational Collaborative Search}

\author{Sandeep Avula}
\authornote{Work done while at the University of North Carolina at Chapel Hill}
\email{sandeavu@amazon.com}
\orcid{0000-0002-8613-4143}
\affiliation{
\institution{Amazon}
\country{USA}
}

\author{Bogeum Choi}
\email{bochoi@unc.edu}
\orcid{0000-0002-9602-233X}
\affiliation{
\institution{University of North Carolina at Chapel Hill}
\country{USA}
}

\author{Jaime Arguello}
\email{jarguello@unc.edu}
\orcid{0000-0002-7645-0556}
\affiliation{
\institution{University of North Carolina at Chapel Hill}
\country{USA}
}

\begin{abstract}
 Our research in this paper lies at the intersection of collaborative and conversational search. We report on a Wizard of Oz lab study in which 27 pairs of participants collaborated on search tasks over the Slack messaging platform. To complete tasks, pairs of collaborators interacted with a so-called \emph{searchbot} with conversational capabilities. The role of the searchbot was played by a reference librarian. It is widely accepted that conversational search systems should be able to engage in \emph{mixed-initiative interaction}---take and relinquish control of a multi-agent conversation as appropriate. Research in discourse analysis differentiates between dialog- and task-level initiative. Taking \emph{dialog-level} initiative involves leading a conversation for the sole purpose of establishing mutual belief between agents. Conversely, taking \emph{task-level} initiative involves leading a conversation with the intent to influence the goals of the other agent(s). Participants in our study experienced three \emph{searchbot conditions}, which varied based on the level of initiative the human searchbot was able to take: (1) no initiative, (2) only dialog-level initiative, and (3) both dialog- and task-level initiative. We investigate the effects of the searchbot condition on six different types of outcomes: (RQ1) perceptions of the searchbot's utility, (RQ2) perceptions of workload, (RQ3) perceptions of the collaboration, (RQ4) patterns of communication and collaboration, and perceived (RQ5) benefits and (RQ6) challenges from engaging with the searchbot.
\end{abstract}



\maketitle

\section{Introduction}
Using search tools to find information has become an integral part of our daily lives. Current search tools come in different forms~\cite{hearst2009search,wilson2011search}. Perhaps the most recognizable one is the Search Engine Results Page (SERP), in which a user enters a keyword query and views a ranked list of results. Within this paradigm, users translate an information need into a keyword query, which the search system uses to find the most relevant results. As successful as this interaction paradigm has been, it is not suitable for supporting \emph{multiple} users collaborating on information-seeking tasks~\cite{morris2008survey,Morris2013}. Addressing this challenge has been the focus of \emph{collaborative search}, a sub-field of \emph{information retrieval} (IR) that focuses on developing systems to support multiple users collaborating on information-seeking tasks.

The most common approach to collaborative search has been to develop standalone systems that include a search interface and peripheral tools for collaborators to communicate, share information, and gain awareness of each other's activities. Prior studies have found that standalone collaborative search systems can provide different benefits~\cite{Htun2017,Morris2007,Putra2018,Shah2013,Capra2012,Paul2009}. However, despite their benefits, standalone systems have not gained widespread adoption. Prior work reports that while people frequently engage in collaborative search, they tend to use \emph{non-integrated} tools (e.g., web search engines and instant messaging platforms) to support these types of collaborations~\cite{Capra2010,morris2008survey,Morris2013}. Morris~\cite{Morris2013} and Hearst~\cite{Hearst2014} discuss these findings as a rationale to integrate search systems into existing communication channels that people \emph{already} use to collaborate, such as messaging platforms. Our research in this paper is a response to this call. Specifically, we explore different ways to integrate search technology into an instant messaging platform (i.e., Slack).

Integrating search tools into a messaging platform raises two important questions. First, what should the search system look like inside a messaging platform? For example, should it follow the same SERP-like paradigm by allowing users to query the system and examine search results directly from the chat channel? Second, what should the system be capable of doing? For example, should it be capable of \emph{proactively} intervening with information that is relevant to the ongoing conversation? Several recent studies in collaborative search have explored these questions~\cite{avula2018searchbots,Avula2019}. Specifically, these studies have investigated the use of \emph{searchbots}---chatbots that can perform specific search operations---to support users during collaborative search~\cite{avula2018searchbots,Avula2019}. Thus far, results have found that searchbots can provide important benefits but also introduce challenges. On one hand, searchbots can help collaborators become more aware of each other's search activities and avoid duplicating effort~\cite{avula2018searchbots,Avula2019}. On the other hand, having collaborators search within the chat channel can be distracting during periods of independent work (i.e., during ``divide and conquer'' activities)~\cite{avula2018searchbots,Avula2019}. Similarly, \emph{proactive} interventions can be either helpful or disruptive, depending on their timing and relevance~\cite{avula2018searchbots}. In this paper, we extend this prior work by investigating the benefits and challenges of \emph{conversational} searchbots during collaborative search.

Within the information retrieval (IR) research community, \emph{conversational search} focuses on developing search systems that can engage with users through dialog-based interaction (either written or spoken)~\cite{Avishek2020,Culpepper2018}. The ultimate goal is to develop systems that can interact with users in the same way that a reference librarian interacts with a library patron, who may have limited domain knowledge and only a vague idea about what information is needed and available to meet their objectives. The leap towards fully conversational search systems is a daunting one. Hence, to provide some guidance, recent research has proposed different capabilities that search systems should have in order to be considered ``conversational''~\cite{Azzopardi2018,Radlinski2017}. Among these is the ability for the system to engage in \emph{mixed-initiative interaction}~\cite{Radlinski2017}. In other words, to be conversational, a search system must be able to take and relinquish control of an information-seeking dialog as appropriate.

Outside of information retrieval, research in discourse analysis has studied mixed-initiative interactions for decades~\cite{Walker1990}. Chu-Carroll and Brown~\cite{chu1997tracking} analyzed mixed-initiative interactions during \emph{task-oriented dialogs}---dialogs between agents with a common objective. Importantly, Chu-Carroll and Brown~\cite{chu1997tracking} argued that initiative during task-oriented dialogs can be defined at two levels: (1) dialog-level initiative and (2) task-level initiative. Both levels of initiative involve taking control of the conversation and placing a discourse obligation on another agent. However, \emph{dialog-level initiative} involves taking control of the conversation in order to better support the current goals of the other agent(s). For example, asking a clarification question in response to a request is a form of taking dialog-level initiative. Conversely, \emph{task-level initiative} involves taking control of the conversation with the intent to \emph{influence} or \emph{alter} the goals of the other agent(s). For example, intervening to provide a suggestion on how to approach the task is a form of taking task-level initiative. With these definitions in mind, an important research question is: From the perspective of users, what are the benefits and challenges associated with virtual assistants (i.e., searchbots) that can take dialog- and task-level initiative to support users during collaborative search?

We report on a Wizard of Oz study in which 27 pairs of participants completed 3 collaborative search tasks over the Slack messaging platform. All three search tasks were \emph{decision-making tasks} that asked participants to consider different alternatives along a given set of dimensions and make a joint selection. To gather information, participants interacted with a searchbot directly from the Slack channel. The role of the searchbot was played by a reference librarian (referred to as the Wizard) from our university. During the study, each participant pair experienced three \emph{searchbot conditions}, which varied based on the level of initiative the searchbot was able to take during the task. In all three conditions, participants were \emph{unable} to search on their own browsers and had to gather information by sending \emph{information requests} to the searchbot directly from the Slack channel. Participants sent information requests to the searchbot (referred to as ``Max'') by using the ``@max'' command (e.g., ``@max What are volunteering opportunities in South Africa?''). In the \textsc{BotInfo} condition, the searchbot could \emph{not} take any form of initiative. The searchbot processed information requests by searching the web and embedding a single search result in the Slack channel. In the \textsc{BotDialog} condition, the searchbot could take dialog-level initiative by asking any number of clarification questions in response to a request. Finally, in the \textsc{BotTask} condition, the searchbot could take both dialog- and task-level initiative. That is, in addition to asking clarification questions, the searchbot could take the initiative by providing task-level suggestions. The searchbot could provide task-level suggestions either in response to a request or by \emph{proactively} intervening in the conversation.

Our study investigated six research questions, which focused on the impact of the searchbot condition on different types of outcomes. Our first three research questions focused on: (RQ1) perceptions of the searchbot's utility, (RQ2) perceptions of workload, and (RQ3) perceptions of the collaborative experience. Our fourth research question (RQ4) focused on communication patterns between participants and the searchbot and the extent to which participants explored different dimensions when comparing alternatives during the task. Finally, our last two research questions (RQ5-RQ6) focused on perceived benefits and challenges associated with the searchbot in a specific condition. To address RQ5-RQ6, we conducted a qualitative analysis of responses to two open-ended questions about the searchbot: How was the searchbot (not) helpful during the task and why?

As researchers push towards fully conversational search systems, it is important to understand the benefits and pitfalls of systems that can take different levels of initiative. Our research focuses on the impact of mixed-initiative interactions in the context of collaborative search, which has not been investigated in prior work. We discuss how our results have implications for designing mixed-initiative conversational search systems in this complex scenario.
\section{Related Work}
Our research builds on four areas of prior work: (1) collaborative search, (2) conversational search, (3) mixed-initiative interactions, and (4) reference services.

\textbf{Collaborative Search:} Collaborative search happens when multiple searchers work together on an information-seeking task. To date, the most prevalent approach to support this practice has been to develop standalone systems~\cite{Capra2012,Golovchinsky2009,Morris2006teamsearch,Morris2007,Morris2013,Paul2009,Putra2018,Shah2013,Yue2012,shah2010coagmento}, which typically include a search engine and peripheral tools for collaborators to communicate, share information, and gain awareness of each other's activities. Prior evaluations of these standalone systems have found that they offer a wide range of benefits. For example, prior studies have found that raising awareness of each other's activities can enable collaborators to learn from each other~\cite{Htun2017,Morris2007}, avoid duplicating work~\cite{Morris2007,Putra2018,Shah2013,Capra2012}, review each other's work~\cite{Htun2017,Capra2012}, delegate tasks~\cite{Paul2009}, and track their progress~\cite{Shah2013}. Despite their benefits, standalone systems have not gained widespread adoption~\cite{Morris2013,Hearst2014}. In a survey by Morris~\cite{Morris2013}, about 50\% of respondents reported doing collaborative searches at least once a week. However, none of the respondents reported using standalone systems specifically designed for collaborative search. Instead, respondents reported using non-integrated tools such as web search engines and communication tools such as phone, email, and instant messaging. Interestingly, while respondents preferred using non-integrated tools that are part of their everyday routines, they acknowledged facing challenges while using these non-integrated tools (e.g., duplicated effort). Based on these results, Morris concluded that future research should investigate ``glue systems'' that can integrate existing search and communication platforms~\cite{Morris2013}.

A few studies have investigated the types of ``glue systems'' advocated by Morris. The SearchBuddies system~\cite{Hecht2012} was developed to automatically embed search results in response to a question posted on social media. Results found opportunities and challenges for ``socially embedded search systems''. For example, on one hand, users responded positively to the system when its results \emph{complemented} those from human users. On the other hand, users responded negatively when the results were relevant but also obvious. Closely related to our work, prior research has also investigated embedding \emph{searchbots} into messaging platforms to support collaborative search~\cite{avula2018searchbots,Avula2019}. \citet{Avula2019} conducted a study in which participants could: (1) only search inside Slack, (2) only search on their individual browsers, and (3) both. Participants reported greater collaborative awareness when they could search inside Slack. However, participants also reported being distracted by their partner's interactions with the searchbot. The authors hypothesized that these distractions probably occurred while participants worked independently on different parts of the task. \citet{avula2018searchbots} conducted a Wizard of Oz study that investigated searchbots that \emph{intervene} in a Slack conversation when they detect an information need. The searchbot was operated by a human. The study considered two types of proactive interventions: (1) eliciting information before providing search results and (2) directly providing search results by ``inferring'' the information need from the conversation. Regardless of the intervention type, participants reported a better collaborative experience when they had access to the searchbot versus a baseline condition where they could only search independently. Additionally, results found that the point of intervention is key. Participants perceived interventions to be disruptive when they were too soon (i.e., before participants understood the task requirements), too late (i.e., after participants had committed to an approach to the task), and when participants were engaged in independent activities.
 
\textbf{Conversational Search:} The goal of conversational search is to develop systems that allow searchers to resolve information needs through dialog-based interaction (written or spoken). In their theoretical framework, Radlinski and Craswell~\cite{Radlinski2017} proposed five capabilities that are desirable for a search system to be considered ``conversational''. Among these is the ability for the system to engage in \emph{mixed-initiative} interaction. During a mixed-initiative conversation, agents take and relinquish control of the conversation to fulfill different objectives~\cite{Walker1990}. Prior research in discourse analysis has argued that goal-oriented dialogs involve two levels of initiative: dialog-level initiative and task-level initiative~\cite{chu1997tracking}. Dialog-level initiative involves taking control of a conversation for the sole purpose of establishing mutual belief between agents. For example, taking dialog-level initiative may involve asking a clarification question in response to a request. Conversely, task-level initiative involves taking control of a conversation with an intent to alter the goals of the other agent(s). For example, taking task-level initiative may involve proposing an alternative approach to the agents' task. Both levels of initiative place a discourse obligation on the other agent(s). However, task-level initiative involves ``directing how the agents' task should be accomplished''.~\cite[p.263]{chu1997tracking}.

Prior work has aimed at defining the \emph{action space} of a conversational search system~\cite{Vakulenko2019,Azzopardi2018}. Some actions place a discourse obligation on the user. For example, \citet{Vakulenko2019} proposed that conversational search systems should be able to elicit information about a user's need and request feedback about available options. These actions are examples of a system taking dialog-level initiative---eliciting information to better understand and address a user's current need. \citet{Azzopardi2018} proposed that systems should also consider and suggest alternatives that have not been discussed, referred to as \emph{alternative information needs}. Such an action would be an example of a system taking task-level initiative---making suggestions to alter a user's objective.

\hl{As previously noted, developing a fully conversational search system is a daunting task.  As a starting point, several studies have aimed at better understanding information-seeking dialogs between \emph{humans}.  Similar to our research, prior studies have used a Wizard of Oz methodology to analyze information-seeking conversations in which one study participant plays the role of the information \emph{seeker} (i.e., the user) and another plays the role of the information \emph{provider} (i.e., the system)~\mbox{\cite{avula2018searchbots,vtyurina2017exploring,trippas2017people,budzianowski2018multiwoz,aliannejadi2019asking}}. These studies have provided insights about desirable capabilities of a conversational search system. Importantly, in all of these studies, the role of the Wizard was played by regular participants or crowdsourced workers. As a methodological contribution, in our study, the role of the Wizard was played by a reference librarian who is formally trained in conducting reference interviews with library patrons. Vakulenko et al.~\mbox{\cite{vakulenko2021large}} analyzed different dialog datasets gathered using a Wizard of Oz methodology. Additionally, the authors included a new dataset of virtual reference interviews between real-world librarians and library patrons. While the study did not distinguish between dialog- and task-level initiative, results found that reference librarians take \emph{high} levels of initiative to support searchers.}

\hl{From a system perspective, research has primarily focused on developing conversational search systems that can take dialog-level initiative by asking clarification questions in response to a request. Prior work has focused on predicting \emph{when} to ask a clarification question~\mbox{\cite{Arguello2017,zhang2020evaluating,christakopoulou2016towards}} and \emph{which} clarification question(s) to ask in a given context~\mbox{\cite{aliannejadi2019asking,hashemi2020guided,rao2018learning,rao2019answer,sun2018conversational,zamani2020analyzing,zamani2020generating}}}.

\textbf{Mixed-Initiative Interactions and Reference Services:} Outside of IR and collaborative search, researchers have investigated the opportunities and challenges of mixed-initiative systems in domains such as human-computer interaction (HCI)~\cite{amershi2019guidelines,horvitz1999principles}, human-robot interaction~\cite{kortenkamp1997traded,jiang2015mixed,chanel2020towards}, and intelligent tutoring~\cite{carbonell1970ai,caine2006mits,graesser2005autotutor,freedman1997degrees}.

Horovitz~\cite{horvitz1999principles} proposed a set of principles for designing mixed-initiative interfaces that can proactively intervene to help users complete tasks. Horovitz proposed that systems should: (1) focus on proactive interventions that have \emph{obvious} value to users; (2) tailor interventions by modeling costs, benefits, and uncertainties about a user's current task and focus of attention; (3) use dialog to reduce key uncertainties; and (4) allow users to easily terminate unhelpful or untimely interventions. Amershi et al.~\cite{amershi2019guidelines} proposed 18 principles for designing AI-infused services. The proposed principles were broadly related to: (1) managing expectations about the system's capacities; (2) predicting when and how to intervene; (3) learning from unhelpful interventions; and (4) communicating to users how the system evolves over time. Additionally, Amershi et al.~\cite{amershi2019guidelines} advocated that systems should enable users to easily understand the \emph{rationale} behind its interventions.

Research has also considered the role of mixed-initiative interaction during human-robot collaborations. Jiang and Arkin~\cite{jiang2015mixed} proposed a taxonomy for classifying mixed-initiative robotic systems along three dimensions: (1) level of proactivity; (2) the extent to which proactive interventions are designed to support goal development, strategy planning, and/or strategy execution; and (3) the mechanisms through which initiative is ``handed off'' from one agent to another. \hl{Studies in this area have also identified challenges related to mixed-initiative robotic systems, such as having humans \emph{misread} the level of initiative a robot intends to take and having too much system initiative lead humans to disengage with the task~\mbox{\cite{chanel2020towards}}}.

Within the area of intelligent tutors, prior work has also argued that mixed-initiative systems can improve learning outcomes~\cite{freedman1997degrees}. Studies suggest that system initiative is more likely to be effective in domains where the learner has low levels of prior knowledge~\cite{graesser2005autotutor} and domains that do not require highly precise terminology~\cite{graesser2005autotutor}. Additionally, research has advocated that systems should avoid eliciting user responses that cannot be fully understood and acted upon by the system~\cite{freedman1997degrees}.

In the field of library science, reference librarians are formally trained in conducting \emph{reference interviews}---mixed-initiative dialogs in which the reference librarian helps a library patron resolve an information need. Similar to the research above, prior work has proposed principles that reference librarians should follow, such as: (1) being transparent about knowledge gaps, (2) treating patrons as equal partners, and (3) approaching each patron as a \emph{unique} case~\cite{mabry2004reference}. Early work by Brooks and Belkin~\cite{Brooks1983} proposed that reference interviews are a useful resource to gain insights into how IR systems can use back-and-forth interactions to learn about users' needs. More recently, Vakulenko et al.~\cite{vakulenko2021large} used virtual reference interviews to better understand the types of actions conversational search systems should be capable of undertaking.
 
Our research in this paper extends prior work in two important ways. First, we investigate the potential benefits and challenges of a conversational search system within the context of collaborative search, bridging two areas of prior research. Second, we investigate the potential benefits and challenges of conversational search systems that can take different levels of initiative (i.e., dialog- and task-level initiative). Current research is mainly focused on developing search systems that only take dialog-level initiative (e.g., ask clarification questions). However, further into the future, we are likely to see conversation search systems that can take task-level initiative (e.g., make task-level suggestions). In Section~\ref{sec:discussion}, we discuss how our results have implications for designing systems that can also take task-level initiative to support collaborative search.

\section{Methods}\label{sec:methods}
To investigate RQ1-RQ6, we conducted a Wizard of Oz laboratory study with 27 pairs of participants (11 male, 33 female, and 10 did not specify). Participants were undergraduate students recruited from our university. We wanted participant pairs to be familiar with each other. Therefore, participants were enrolled in the study in pairs (i.e., were friends or acquaintances). This recruitment technique has been used in prior collaborative search studies~\cite{avula2018searchbots,Avula2019}. The role of the searchbot was played by a reference librarian (referred to as the ``Wizard'') from one of the libraries at our university. Three reference librarians (all female) participated in the study.

Each pair of participants worked on three collaborative search tasks and interacted with the searchbot to gather information. Participants experienced three searchbot conditions (Section~\ref{sec:conditions}), which varied based on the level of initiative the searchbot could take. We were \emph{not} interested in investigating task effects. Therefore, to keep the search tasks as consistent as possible, we designed three tasks that asked participants to compare different alternatives along a given set of criteria and make a joint selection (Section~\ref{sec:tasks}). Participants were seated in different rooms and communicated with each other (and the searchbot) over Slack. Participants were unable to search on their own and had to gather information by issuing \emph{information requests} to the searchbot directly from Slack. The Wizard, playing the role of the searchbot, sat in a third room and had access to two resources to assist the participants. First, they had access to the participants' Slack channel in order to monitor the conversation. Second, they used a custom-built web application to: (1) search the web, (2) forward individual search results to the participants' Slack channel, and (3) send messages to the participants through Slack (i.e., clarification questions or suggestions, depending on the searchbot condition). Participants could open search results in their own browsers, but were instructed to not use the browser to search on their own. Across all three conditions, the Wizard used the same search system, developed using Bing's web search API.
\subsection{Study Protocol and Design}\label{sec:protocol}

\begin{figure*}[t]
	\centering
	\begin{subfigure}[t]{0.45\textwidth}
		\centering
		\includegraphics[width =\columnwidth]{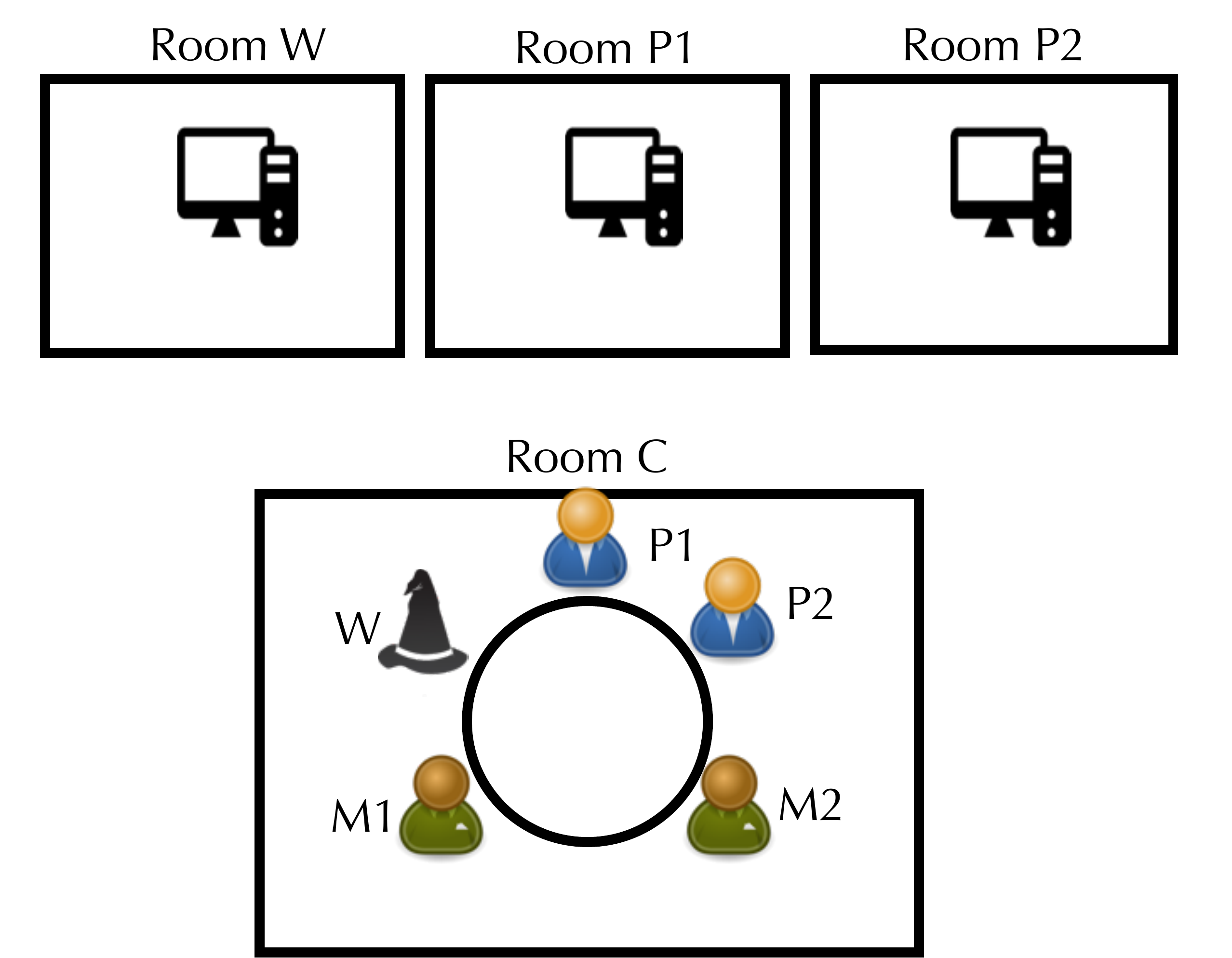}
		\caption{At the start of each study session, the lead moderator (M1) explained the purpose and protocol of the study to both participants (P1 and P2). Additionally, before each task, the lead moderator explained the next searchbot condition in the \emph{presence} of the Wizard (W). Participants were allowed to ask questions about the searchbot's capabilities in the next condition.}
		\label{subfig:intro}
	\end{subfigure}
	\hspace{1cm}
	\begin{subfigure}[t]{0.45\textwidth}
		\centering
		\includegraphics[width =\columnwidth]{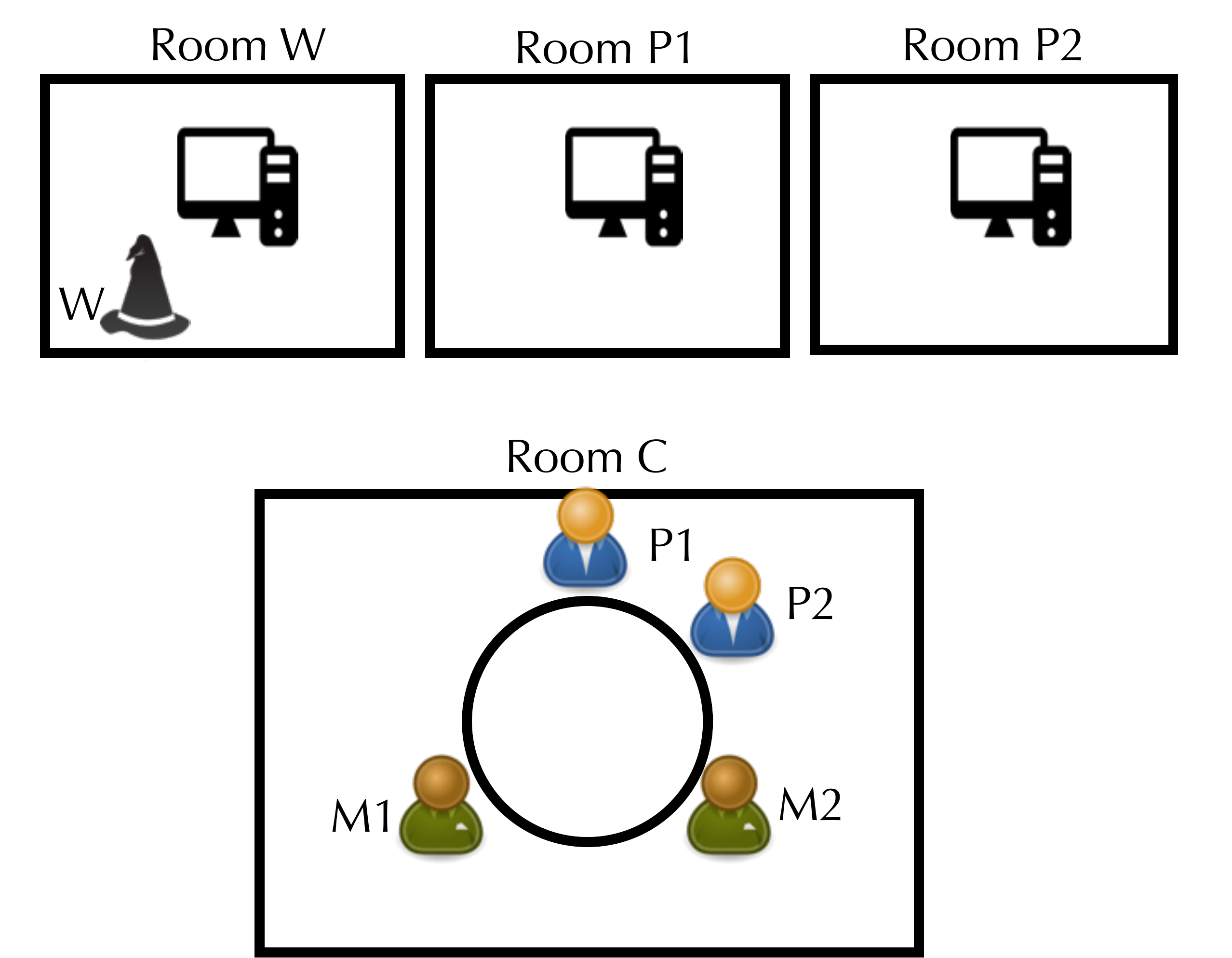}
		\caption{After explaining the next searchbot condition, the lead moderator read the next search task description aloud to the participants in the \emph{absence} of the Wizard. The Wizard was absent because we did not want them to learn about the specific criteria participants were asked to consider during the task.}
		\label{subfig:task}
	\end{subfigure}	
	\begin{subfigure}[t]{0.45\textwidth}
		\centering
		\includegraphics[width =\columnwidth]{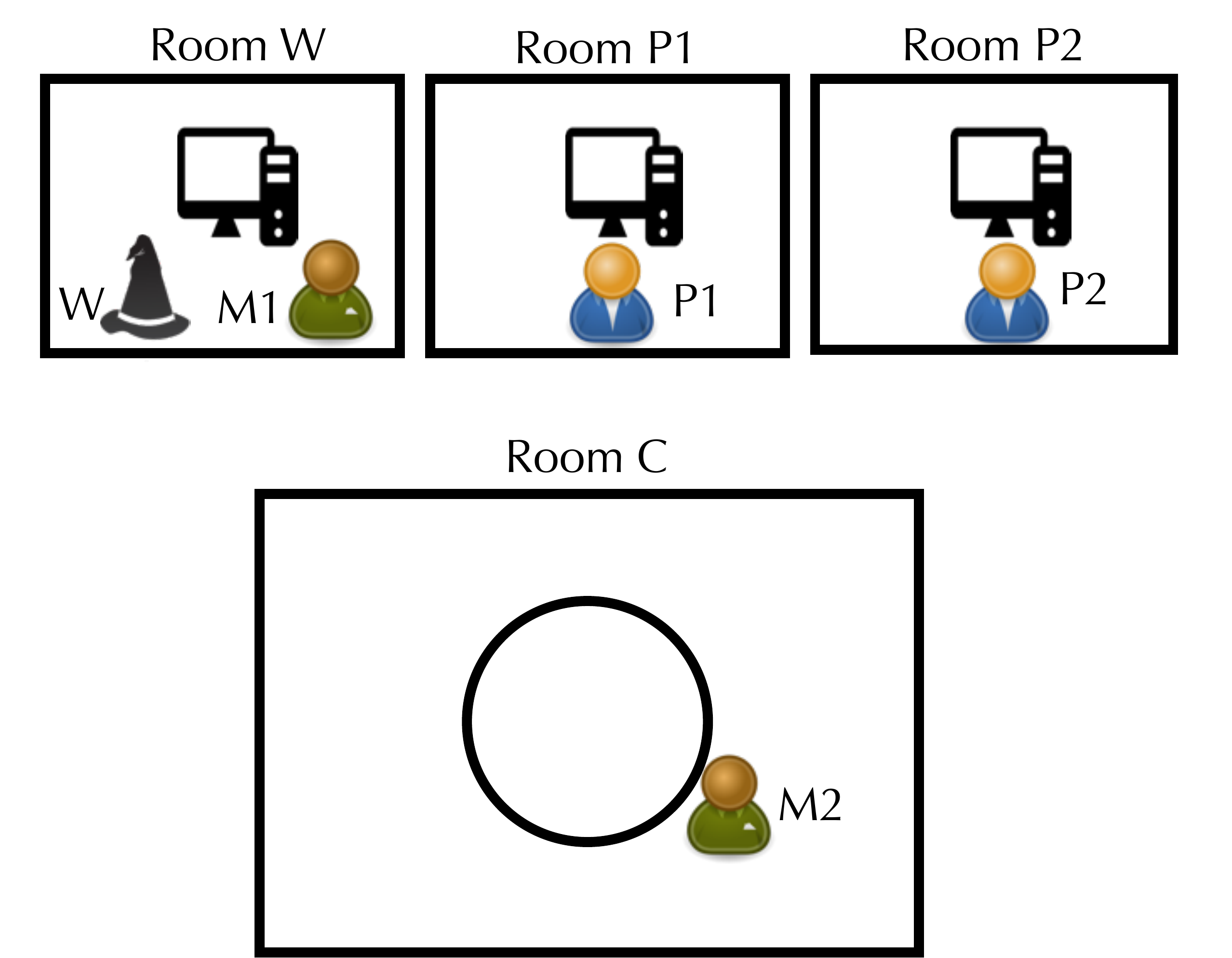}
		\caption{During each search task, both participants and the Wizard sat in separate rooms. The lead moderator sat behind the Wizard to assist with any technical difficulties.}
		\label{subfig:search}
	\end{subfigure}
	\hspace{1cm}
	\begin{subfigure}[t]{0.45\textwidth}
		\centering
		\includegraphics[width =\columnwidth]{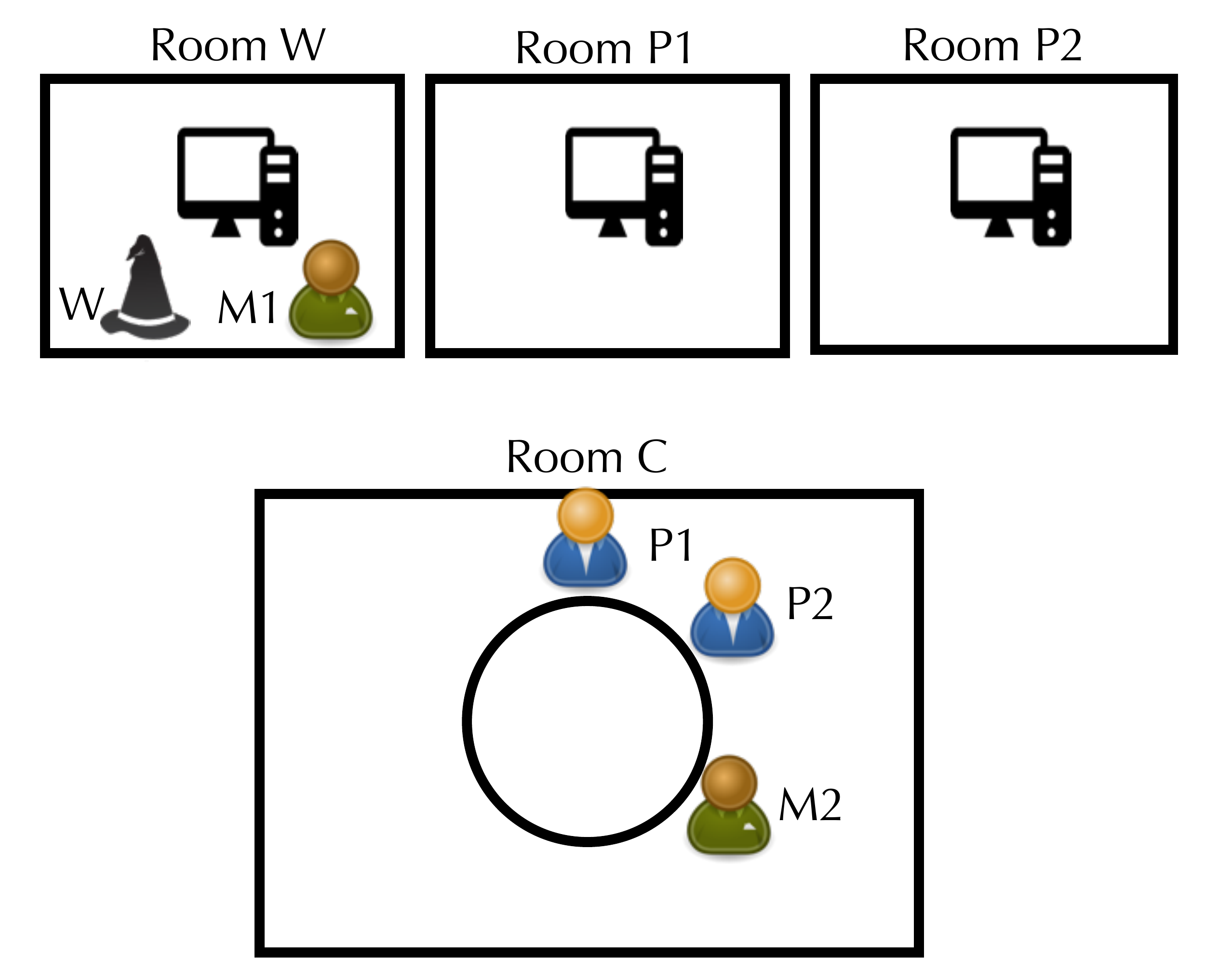}
		\caption{After each task, both participants were asked to verbally explain their solution to the task to the secondary moderator (M2). This exercise was done to discourage participants from satisficing.}
		\label{subfig:solution}
	\end{subfigure}
	\caption{Physical setup during different phases of the study session.}\label{fig:rooms}
\end{figure*}

The study involved two moderators and took place in a physical space with four rooms: one common room (\textbf{C}), one room for the Wizard (\textbf{W}), and one room for each participant (\textbf{P1} and \textbf{P2}). Figure~\ref{fig:rooms} illustrates where members of the study were located during different phases of the study. 

\textbf{Study Protocol:} Each study session proceeded as follows. At the start of each session, both moderators welcomed the participants and the Wizard in room \textbf{C} (Figure~\ref{subfig:intro}). Here, the lead moderator explained the purpose and protocol of the study. 

Next, participants completed three tasks that followed the same sequence of steps. First, the lead moderator explained the next searchbot condition to the participants (Figure~\ref{subfig:intro}). The lead moderator described the searchbot's capabilities in the presence of the Wizard so that participants could ask any clarifying questions directly to the Wizard. Next, the Wizard left for room \textbf{W} and the lead moderator read the next task description aloud to the participants (Figure~\ref{subfig:task}). Participants were also given printed copies of the task description for reference. The Wizard was absent when participants were described the task because we did not want the Wizard to become aware of any task-specific objectives. In other words, we wanted to simulate a scenario in which the ``system'' is unaware of all the details related to a searcher's objective. After learning about the next task, participants left for their respective rooms \textbf{P1} and \textbf{P2} (Figure~\ref{subfig:search}). Participants were given 15 minutes to complete each task. Participants completed a series of post-task questionnaires after each task. During each task, the lead moderator sat behind the Wizard in room \textbf{W} to assist with any technical issues. After completing each task (i.e., search task + post-task questionnaires), participants met in room \textbf{C} and explained their solution to the task to the secondary moderator (Figure~\ref{subfig:solution}). This exercise was done to discourage participants from satisficing. Urgo et al.~\cite{Urgo2020} used a similar technique to discourage participants from satisficing during learning-oriented search tasks. Each participant received US\$20 for participating and the Wizard received US\$30 per study session.

\textbf{Study Design:} Every pair of participants was exposed to all three searchbot conditions and all three task domains (i.e., a within-subjects design). To control for learning and fatigue effects, we varied the order in which participants were exposed to our three searchbot conditions and task domains. A Latin square for three treatment conditions yields three treatment orders (i.e., ABC, CAB, BCA), which ensures that each treatment appears exactly once in each position. To accommodate our two factors (i.e., searchbot condition and task domain), we used the cross product of two Latin square orderings, which yielded 9 orderings (i.e., 3 searchbot condition orders $\times$ 3 task domain orders = 9 searchbot-task condition orders). Each of these nine sequences was completed by three pairs of participants (i.e., 9 orders $\times$ 3 participant pairs per order = 27 study sessions). Each order was completed by a different Wizard, and each Wizard completed 9 sessions.

\subsection{Searchbot Conditions}\label{sec:conditions}
\begin{figure*}[t]
	\centering
	\begin{subfigure}[c]{0.48\textwidth}
		\centering
		\includegraphics[width =\columnwidth]{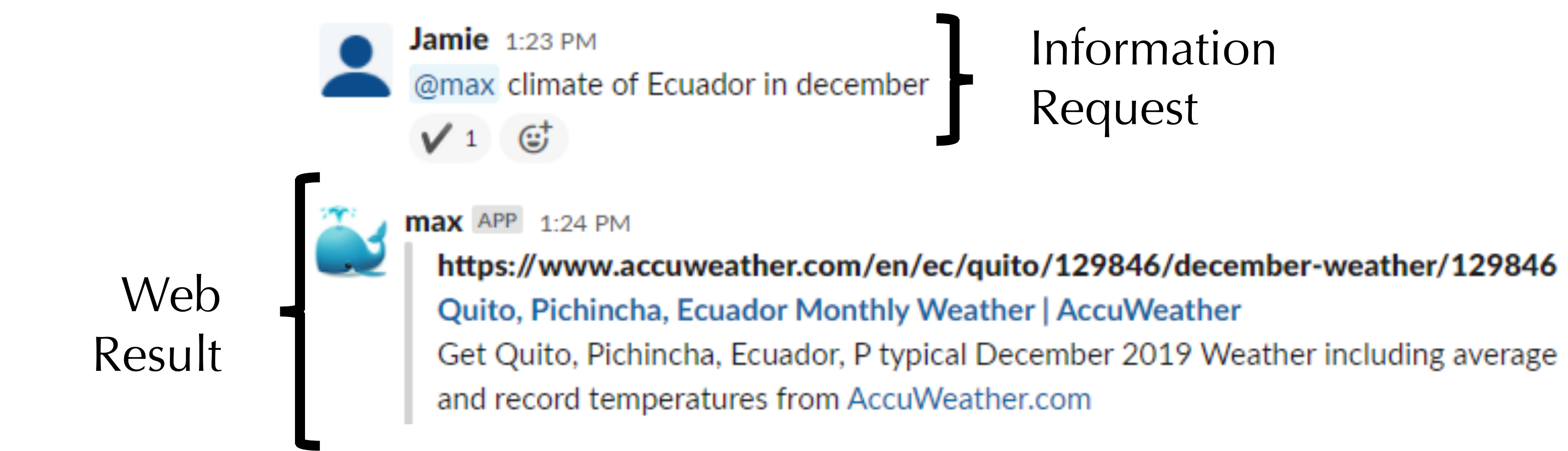}
		\caption{\textsc{BotInfo} Condition. In this example, the searchbot receives a request and returns a web result.}
		\label{subfig:bot_info}
	\end{subfigure}
	\begin{subfigure}[c]{0.48\textwidth}
		\centering
		\includegraphics[width =\columnwidth]{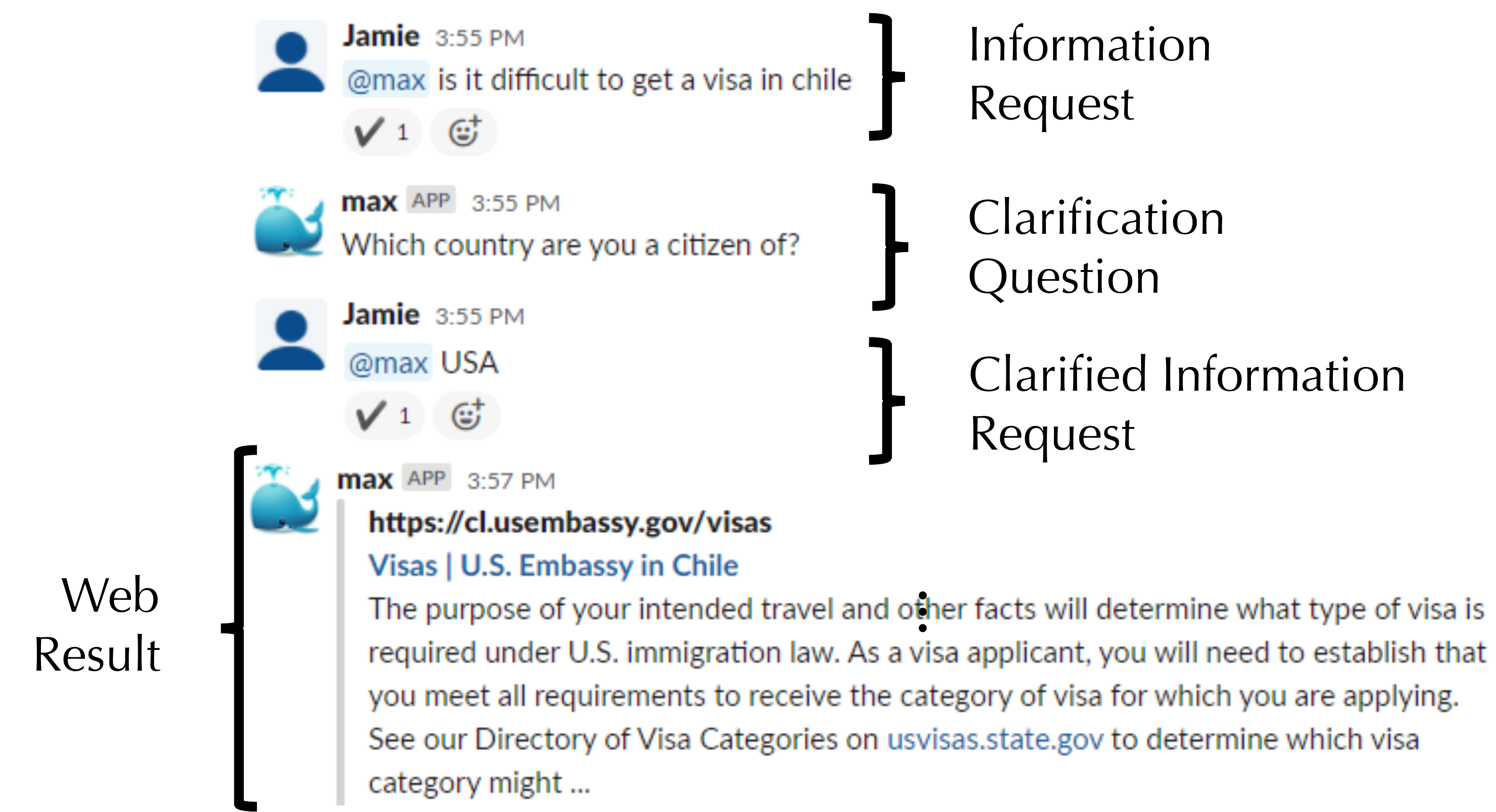}
		\caption{\textsc{BotDialog} Condition. In this example, the searchbot receives an underspecified request and asks for clarification before returning a web result.}
		\label{subfig:bot_dialog}
	\end{subfigure}	
	\begin{subfigure}[c]{1\textwidth}
		\centering
		\includegraphics[width =\columnwidth]{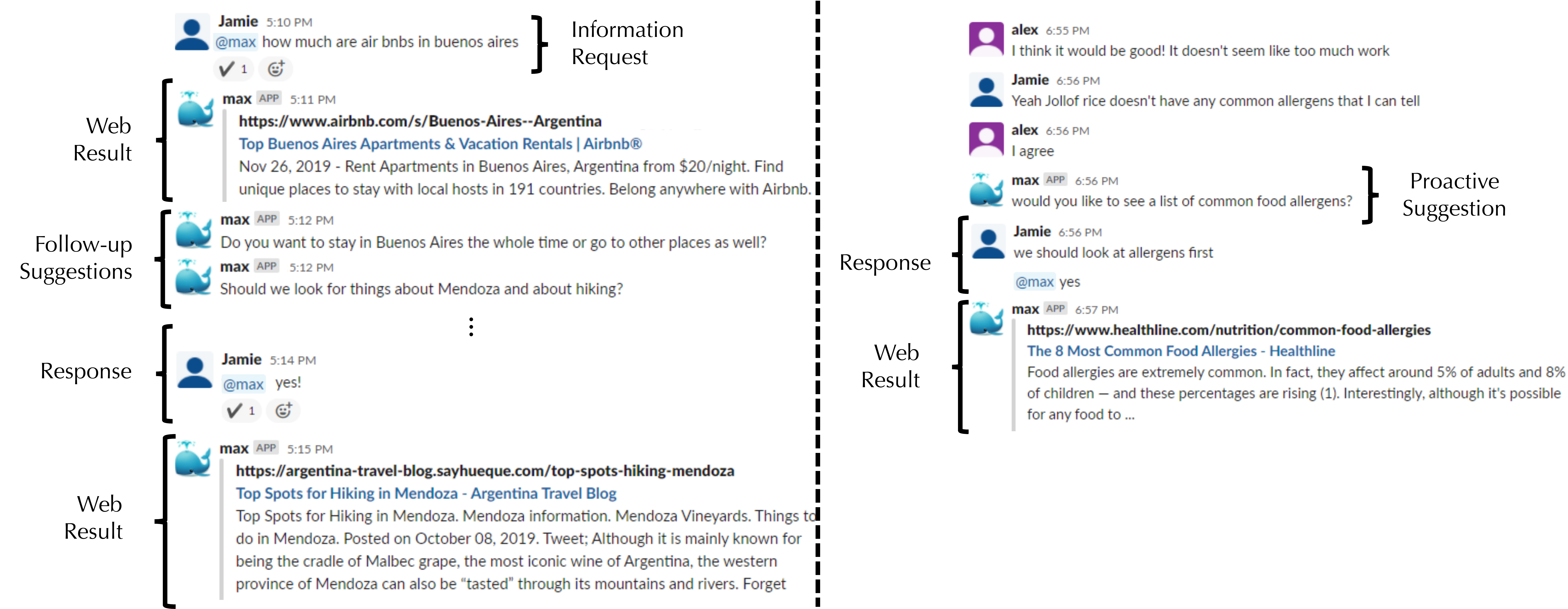}
		\caption{\textsc{BotTask} Condition. In the example on the LEFT, after addressing a request, the searchbot makes a task-level suggestion for participants to consider other cities in Argentina and related activities. In the example on the RIGHT, the searchbot decides that the participants might benefit from knowing about common allergens and makes a proactive suggestion for participants to consider this information at this point in the task.}
		\label{subfig:bot_task}
	\end{subfigure}
	\vspace{-.5cm}
	\caption{Example of searchbot interactions across conditions.\vspace{-.5cm}}\label{fig:searchbot_conditions}
\end{figure*}

Each pair of participants experienced three searchbot conditions (i.e., a within-subjects design). \citet{chu1997tracking} proposed that mixed-initiative, goal-oriented dialogs involve two levels of initiative: dialog-level and task-level initiative. In our three searchbot conditions, the searchbot was able to take different levels of initiative: no initiative, only dialog-level initiative, and both dialog- and task-level initiative. Figure~\ref{fig:searchbot_conditions} shows examples of how the searchbot interacted with the participants in each condition.

\textbf{\textsc{BotInfo} (Figure~\ref{subfig:bot_info}):} In this condition, the Wizard played the role of a searchbot that can only process information requests by responding with a search result. Participants could issue information requests using the `@max' preamble (e.g., ``@max What are the best beaches in South America?''). In response to a request, the Wizard could search the web using our custom-built system and send a single search result by clicking a `send' button positioned next to the result on the search results page (SERP). Clicking this button embedded the search result directly on the participants' Slack channel. The Wizard was instructed not to provide direct answers and only respond with one search result per information request. The \textsc{BotInfo} condition was meant to mimic the current capabilities of systems such as Alexa, Cortana, Google Now, and Siri, which can respond to natural language requests but do not (typically) engage in further dialog.

\textbf{\textsc{BotDialog} (Figure~\ref{subfig:bot_dialog}):} In this condition, the searchbot was able to take dialog-level initiative as characterized by \citet{chu1997tracking}. In a mixed-initiative conversation, an agent takes dialog-level initiative when they lead the conversation for the sole purpose of \emph{establishing mutual belief between agents}. Specifically, in this condition, the Wizard was able to ask one or more clarification questions in response to an information request. The Wizards were instructed to ask clarification questions to ensure a shared understanding of the request. As in all searchbot conditions, participants could issue information requests using the `@max' preamble. In response to a request, the Wizard could ask one or more clarification questions in order to better understand the participants' current need. After asking zero, one, or more clarification questions, the Wizard could send a single search result to the participants' Slack channel. After sending a search result, the Wizard could not ask any additional clarification questions about the request. In other words, any additional clarification questions had to be asked in response to a \emph{new} request. The Wizards were instructed to ask clarification questions when they believed that additional information would enable them to find a better search result and satisfy the participants' current need. In practice, many of these clarification questions were asked when the participants issued information requests that were ambiguous, too broad, or used subjective qualifiers. For example, in response to subjective qualifiers such as `cheap', `warm', or `easy', the Wizard might have asked participants to specify a price range, temperature range, or specific techniques to avoid. The \textsc{BotDialog} condition was meant to mimic the types of conversational search systems we may see in the coming years. Current research is primarily focused on developing conversational search systems that can ask follow-up questions to better address a user's information need~\cite{christakopoulou2016towards,zhang2018towards,aliannejadi2019asking, sun2018conversational}.

\textbf{\textsc{BotTask} (Figure~\ref{subfig:bot_task}):} In this condition, the Wizard was able to take both dialog- and task-level initiative. As in the \textsc{BotDialog} condition, the Wizard was able to ask one or more clarification questions in response to an information request before sending a web result to the participants. Additionally, the Wizard was able to provide task-level suggestions. In other words, following the characterization of task-level initiative from \citet{chu1997tracking}, the Wizard was able to lead the conversation by providing suggestions with the intent to \emph{influence participants' goals or approach to the task}. In this condition, the Wizard was able to provide suggestions either in response to a request or by 
\emph{proactively intervening} in the conversation without being asked. The Wizards monitored the participants' Slack channel and were instructed to intervene if they felt they could provide valuable task-level advice. The \textsc{BotTask} condition was meant to mimic conversational search systems we may see farther into the future, which may be able to proactively influence the goals and strategies of users during a search task.

\subsection{Tasks}\label{sec:tasks}
Participants completed three collaborative search tasks that required them to consider different alternatives along a given set of criteria and make a joint selection. The alternatives were left open-ended. However, the decision criteria were specified in the task description. Each task involved three criteria. We believe that our tasks encouraged participants to engage with each other and the searchbot. During each task, collaborators had to explore viable alternatives, compare them across the given criteria, discuss their individual preferences, converge on a final selection, and develop a logical justification. Participants were given 15 minutes to complete each task. To discourage participants from satisficing, after each task's post-task questionnaires, participants were asked to explain and justify their final selection to the secondary moderator (Figure~\ref{subfig:solution}). Our three tasks had the following themes: (1) volunteer planning, (2) vacation planning, and (3) dinner planning. Each task included a background scenario to contextualize the task and an objective statement. Participants were assigned gender-neutral names (Jamie and Alex) to help them internalize the task scenarios during the study session. To illustrate, the volunteer planning task had the following background and objective.

\textbf{Background:} Jamie and Alex are rising seniors. They have decided that before they finish school, they would like to spend a summer volunteering (2-3 months). Jamie heard from a friend that volunteering internationally is an option they could consider. They have both decided to explore different volunteering programs in Africa.

\textbf{Objective:} Jamie mentioned to Alex that their parents would only agree to a volunteering program in Africa if they are confident of their \emph{safety}, \emph{affordability of the plan}, \emph{and if the project is exciting}. In this task, with the help of the searchbot (i.e., Max), work together to find a project which you both think is most suitable for you.

The three criteria associated with each task were as follows: (1) volunteer planning---(a) safety, (b) affordability, and (c) excitement; (2) vacation planning---(a) things to do/see, (b) ease of obtaining a visa, and (c) safety; and (3) dinner planning---(a) choice of cuisine, (b) difficulty level, and (c) possible allergens to avoid.

\subsection{Post-task questionnaires}\label{sec:questionnaires}
Participants completed three questionnaires after each task. First, participants completed an 8-item questionnaire about their perceptions of the searchbot. Participants were asked about the extent to which the searchbot: (1) found useful information, (2) found everything that was needed, (3) showed readiness to help, (4) showed interest in the participants' needs, (5) understood the participants' needs, (6) provided a satisfying experience, (7) verified the usefulness of information provided, and (8) was disruptive. Items 1-7 were adapted from an existing questionnaire developed to evaluate reference interviews by librarians~\cite{gross2001wants}. Participants responded to agreement statements on a 7-point scale from ``Strongly Disagree'' to ``Strongly Agree''. Participants were also asked two open-ended questions about their perceived benefits and challenges from engaging with the searchbot.

Second, participants completed a 6-item questionnaire about their perceptions of workload~\cite{Hart2006}: (1) mental demand, (2) physical demand, (3) temporal demand, (4) task failure, (5) effort, and (6) frustration. Participants responded to statements on a 7-point scale with labeled endpoints. The item for failure had the endpoints of ``Perfect'' and ``Failure''. The remaining items had the endpoints of ``Very Low'' to ``Very High''. In all cases, higher values indicate higher perceptions of workload.

Third, participants completed an 9-item questionnaire about their perceptions of the collaborative experience: (1) me being aware of my partner's activities, (2) my partner being aware of my activities, (3) maintaining joint attention (i.e., looking at or talking about the same information), (4) ease of sharing information, (5) ease of coordinating, (6) ease of reaching consensus, (7) having a smooth ``flow'' of communication, (8) self enjoyment, and (9) perceptions of my partner's enjoyment. Participants responded to agreement statements on a 7-point scale from ``Strongly Disagree'' to ``Strongly Agree''. All questionnaires are available at: \url{https://bit.ly/3ehM2q7}.

\subsection{Wizard Training Session}\label{sec:wizard_training}
Prior to the study, all three reference librarians attended a Wizard training session. During this training session, the reference librarians were explained the purpose of the study and the three searchbot conditions. We also introduced them to the tools they would use during the study and the three search tasks. We wanted to simulate a realistic scenario in which the search ``system'' is unaware of all the task-specific objectives searchers are trying to accomplish. Thus, search tasks were described in general terms (e.g., ``You will help participants plan a vacation trip to South America.''). For each task, the Wizards were asked to brainstorm together and list topics to consider when asking clarification questions (in the \textsc{BotDialog} and \textsc{BotTask} conditions) and making suggestions (in the \textsc{BotTask} condition). During the study, the Wizards were given printed copies of these lists to refresh their memory about ways to support participants.

\subsection{Communication and Collaboration Measures}\label{sec:interactions}
In RQ4, we investigate whether the searchbot condition influenced the communication patterns amongst participants and between participants and the searchbot. We computed four measures: (1) TME: \underline{total} number of \underline{messages} \underline{exchanged} in the Slack channel; (2) TLS: \underline{total} number of web \underline{links} \underline{sent} by the searchbot to the participants; (3) TEI: \underline{total} number of times participants \underline{explicitly} \underline{interacted} with the searchbot by using the `@max' preamble; and (4) TSI: \underline{total} number of times the \underline{searchbot} took the \underline{initiative} by either asking a clarification question (in the \textsc{BotDialog} and \textsc{BotTask} conditions) or by intervening with a suggestion (in the \textsc{BotTask} condition). The TSI measure was only computed for the \textsc{BotDialog} and \textsc{BotTask} conditions. 

In addition to these four communication measures, we also analyzed the number of \underline{task-relevant} \underline{dimensions} (TRD) considered by participants during the task. As described in Section~\ref{sec:tasks}, during each task, participants were asked to compare different alternatives along three specific dimensions (e.g., compare dishes to cook for a dinner party based on the type of cuisine, difficulty level, and common allergens to avoid). A preliminary analysis found that participants often considered \emph{additional} dimensions beyond those specified in the task description (e.g., novelty, equipment involved, accessibility of ingredients, etc.). Section~\ref{subsec:data_analysis} describes how we measured the number of dimensions considered by participants during a task.

\subsection{Data Analysis}\label{subsec:data_analysis}
To address RQ1-RQ4, we conducted a quantitative analysis of participants' post-task questionnaire responses and patterns of communication and collaboration. To address RQ5-RQ6, we conducted a qualitative analysis of participants' responses to our two open-ended questions about their perceived benefits and challenges from engaging with the searchbot.

\textbf{Quantitative analysis:} For RQ1-RQ4, we used mixed-effects regression models to investigate the main effect of the searchbot condition on different outcomes. Using mixed-effects models enabled us to account for \emph{random} variations across participant pairs ($N$=27) and individual participants ($N$=54). For outcomes at the participant-pair level (e.g., total messages exchanged), we included the \emph{participant-pair ID} as a random factor. Conversely, for outcomes at the participant level (e.g., effort), we used nested random effects---the \emph{participant ID} was nested with the \emph{participant-pair ID}. Additionally, we included four fixed factors in our models: searchbot condition, task ID, wizard ID, and wizard-session ID. In RQ1-RQ4, we investigate the effects of the searchbot condition on different types of outcomes. Therefore, searchbot condition was included as the main independent variable in our RQ1-RQ4 models. Each participant-pair completed three tasks with the same wizard (Section~\ref{sec:tasks}), and each wizard completed 9 (out of 27) study sessions. To control for differences at the task and wizard level (e.g., some tasks being more difficult or some wizards being more effective), we included task ID and wizard ID as \emph{control} variables. Finally, we wanted to account for the possibility of wizards becoming more effective as they completed more study sessions. To this end, we also included the wizard-session ID (i.e., values 1-9) as a \emph{control} variable. To test the statistical significance of each model, we computed the $\chi^2$ statistic using the likelihood-ratio test against a null model (i.e., one without the searchbot condition as a covariate). To compare between all pairs of searchbot conditions, we ran analyses using both the \textsc{BotInfo} and \textsc{BotDialog} conditions as baselines.

\textbf{Qualitative analysis:} As part of RQ4, we measured the number of distinct dimensions participants considered during each task. To compute this measure, we conducted a content analysis of participants' chat logs. First, to test the reliability of this coding effort, two of the authors independently coded chat logs from three randomly selected participant-pairs (3/27 = 11\% of the data). This resulted in an average Jaccard coefficient of 91\%.\footnote{The Jaccard coefficient measures the similarity between two sets of items---the intersection divided by the union. Therefore, out of all the dimensions identified by \emph{either} author (i.e., the union), 91\% were identified by \emph{both} authors (i.e., the intersection).} Given this high level of agreement, one author coded the remaining chat logs.

To investigate RQ5-RQ6, we conducted an inductive content analysis of participants' responses to our two open-ended post-task questions about their perceived benefits and challenges from engaging with the searchbot in a specific condition. The coding process proceeded as follows. First, two of the authors independently coded a third of the data and developed an initial set of codes. Then, both authors met to discuss and merge their individual codes (i.e., names and definitions). Next, both authors independently coded another third of the data by applying the existing codes and creating new ones as needed. After this, the authors met to finalize the codebook. Finally, both authors independently (re-)coded 100\% of the data and then met to discuss their assigned codes and reconcile any differences. Our qualitative codes were not mutually exclusive---responses could be associated with multiple codes.

\subsection{Decisions and Rationale}\label{subsec:rationales}

Our study design involves several decisions that deserve additional explanation.

\textbf{Participants Knew the Searchbot was Human:} In a traditional ``Wizard of Oz'' study, participants interact with a ``computer system'' that is unknowingly operated (partly or completely) by an unseen human~\cite{Martin2012}. In our study, participants were fully aware that the role of the searchbot was being played by a human (i.e., a reference librarian). Regardless of the searchbot condition, participants knew there was a human behind the scenes. This decision was made for two reasons. 

First, prior work has found that \emph{expectations} strongly influence users' perceptions of an AI system (e.g., usability and willingness to collaborate with the system)~\cite{Khadpe2020,Luger2016}. Therefore, we wanted participants to have \emph{consistent} expectations about the searchbot's ``intelligence'', both across searchbot conditions and across study sessions. For example, we did not want participants to think of the searchbot as being more ``intelligent'' in the \textsc{BotTask} versus the \textsc{BotInfo} condition. Similarly, we did not want participants to have different expectations based on their individual assumptions about what a fully automated system should be capable of doing.

Second, an important goal of the study was to investigate the influences (e.g., perceived benefits and challenges) of a conversational agent that can provide task-level advice. Providing task-level advice (and doing so \emph{proactively}) goes well beyond the capabilities of current commercial systems such as Alexa, Cortana, and Siri. Thus, we found it unlikely for participants to believe they were interacting with a fully automated system (i.e., no ``human in the loop'') in the \textsc{BotTask} condition.

\textbf{Participants Knew About the Searchbot's Capabilities Before Each Task:} The primary moderator explained the searchbot's capabilities before each searchbot condition (Figure~\ref{subfig:intro}). Additionally, this explanation took place in front of the Wizard so that participants could ask any clarifying questions. This was done to set expectations from the outset of the task. Participants were given 15 minutes to complete each task. Thus, we did not want participants to spend any part of this time trying to determine the capabilities of the searchbot through trial-and-error.

\textbf{The Searchbot Returned a Single Web Result:} In all three searchbot conditions, the searchbot responded to information requests by embedding a \emph{single} search result (vs.~a ranked list of results) directly in the participants' Slack channel. The decision to return a single result was made to encourage participants to engage with the searchbot throughout the search session. We believe that providing multiple search results at a time would have significantly reduced participants' engagement with the searchbot.

\section{Results}

\subsection{RQ1: Searchbot Utility}

In RQ1, we investigate the effects of the searchbot condition on participants' post-task perceptions of the searchbot's utility: (1) usefulness, (2) coverage, (3) readiness, (4) interest, (5) understanding, (6) satisfaction, (7) verification, and (8) disruption. The questionnaire items about interest and verification were only included in the \textsc{BotDialog} and \textsc{BotTask} conditions, as those were the only conditions in which the searchbot could exhibit these behaviors. 

\begin{figure}[t]
	\centering
	\includegraphics[width=1\textwidth]{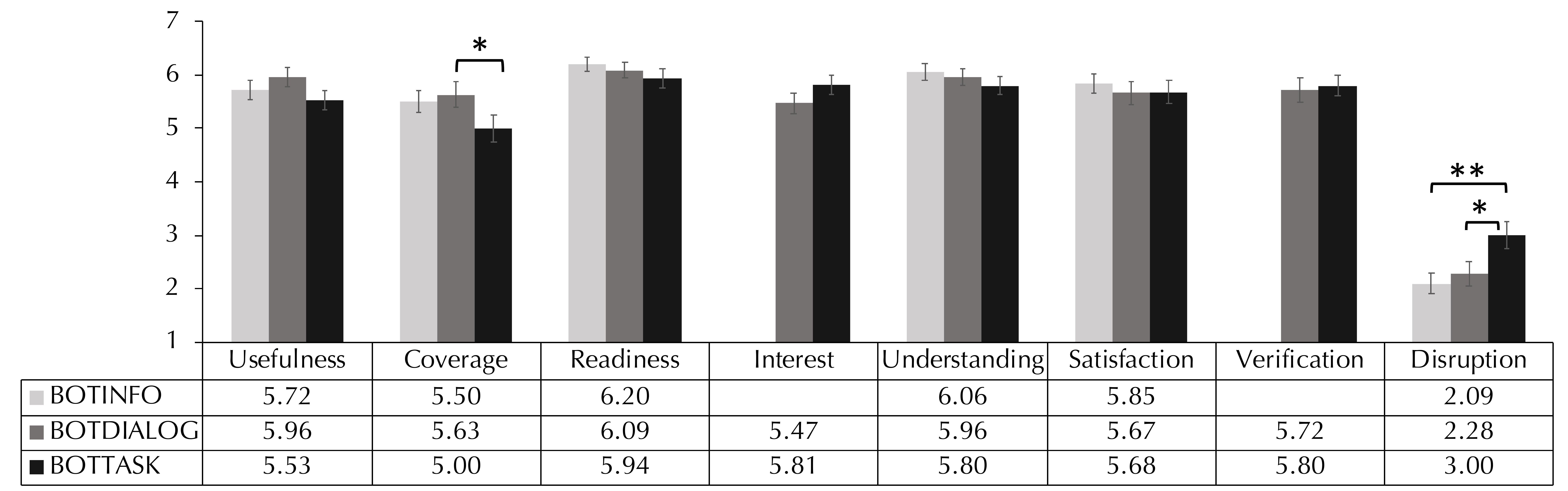}
	\caption{RQ1: Effects of the searchbot condition on participants' perceptions of the searchbot's utility. Symbols `*' and `**' denote significant differences at \emph{p} < .05, and \emph{p} < .01 level.\vspace{-.5cm}}\label{fig:utility}
\end{figure}

Figure~\ref{fig:utility} shows the means and 95\% confidence intervals across searchbot conditions. The searchbot condition had a significant effect for disruption (${\chi(2)}^2$=11.731, $p$<.01) and a marginally significant effect for coverage (${\chi(2)}^2$=5.551, $p$=.06). In terms of disruption, participants reported being significantly more disrupted by the searchbot in the \textsc{BotTask} versus \textsc{BotInfo} ($\beta$=0.920, S.E.=0.279, $p$<.01) and \textsc{BotDialog} condition ($\beta$=0.720, S.E.=0.277, $p$<.05). In terms of coverage, participants reported covering less task-relevant information in the \textsc{BotTask} versus \textsc{BotDialog} condition ($\beta$=-0.630, S.E.=0.280, $p$<.05).

\begin{figure}[t]
	\centering
	\hspace*{-.2in}
	\includegraphics[width=0.70\textwidth]{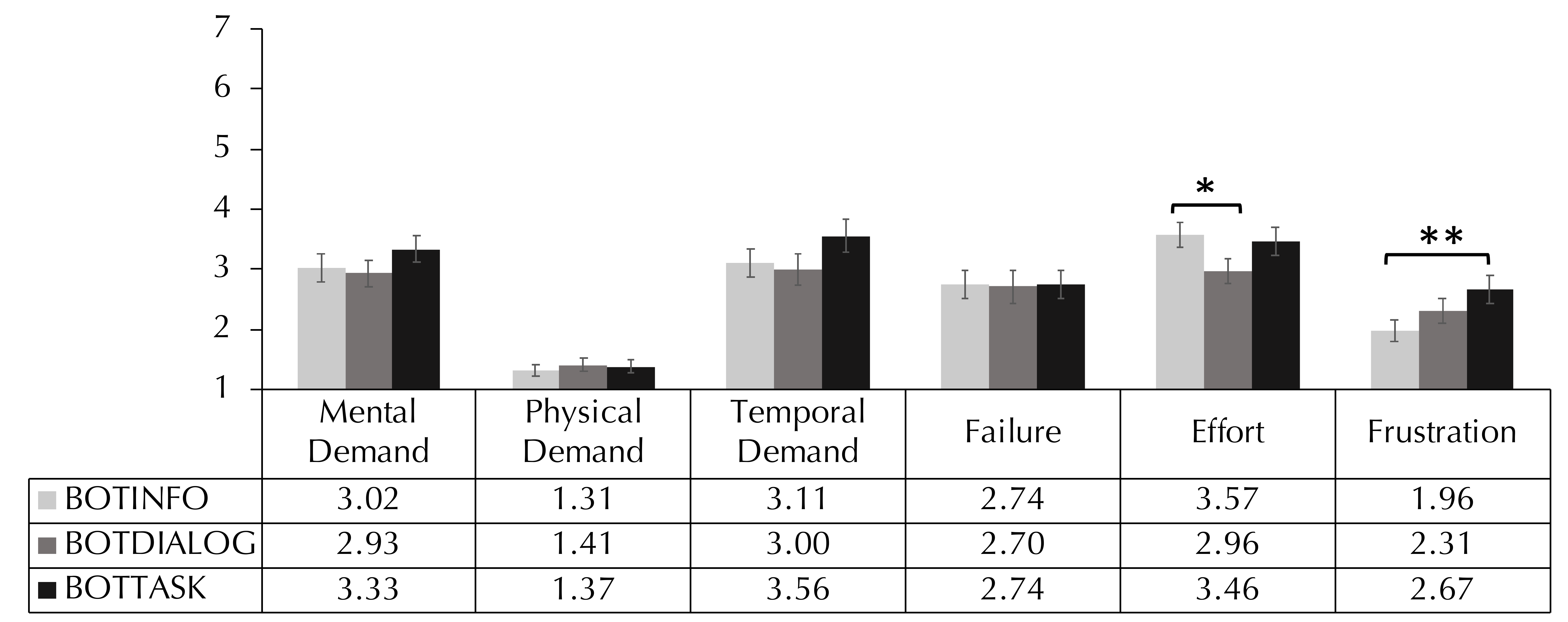}
	\caption{RQ2: Effects of the searchbot condition on participants' perceptions of workload. Symbols `*' and `**' denote significant differences at \emph{p} < .05, and \emph{p} < .01 level.\vspace{-.5cm}}\label{fig:workload}
\end{figure}

\subsection{RQ2: Workload}
In RQ2, we investigate the effects of the searchbot condition on participants' post-task perceptions of workload: (1) mental demand, (2) physical demand, (3) temporal demand, (4) failure, (5) effort, and (6) frustration. Figure~\ref{fig:workload} shows the means and 95\% confidence intervals across the searchbot conditions. The searchbot condition had a significant effect for frustration (${\chi(2)}^2$=10.476, $p$<.01) and effort (${\chi(2)}^2$=5.997, $p$<.05). In terms of frustration, participants reported significantly greater frustration in the \textsc{BotTask} versus \textsc{BotInfo} condition ($\beta$=0.704, S.E.=0.212, $p$<.01). In terms of effort, participants reported significantly greater effort in the \textsc{BotInfo} versus \textsc{BotDialog} condition ($\beta$=0.611, S.E.=0.263, $p$<.05). Participants also reported greater effort in the \textsc{BotTask} versus \textsc{BotDialog} condition. However, this effect was only marginally significant ($\beta$=0.5, S.E.=0.263, $p$=.06).

\subsection{RQ3: Collaborative Experience} 

\begin{figure}[t]
	\hspace{-.8cm}
	\centering
	\includegraphics[width=1.05\textwidth]{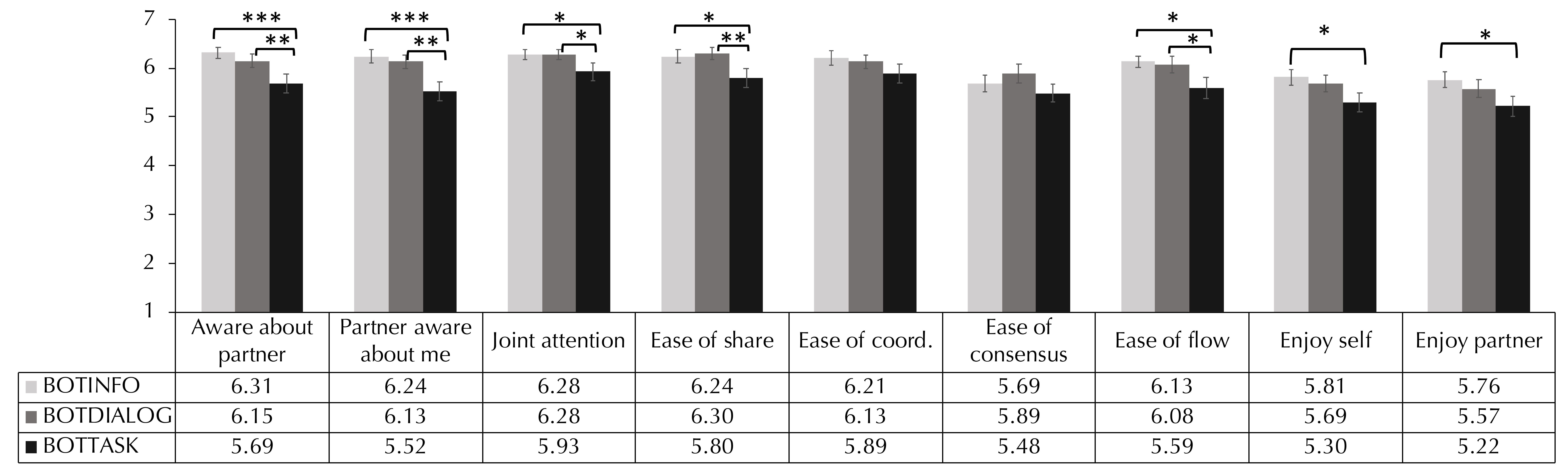}
	\caption{RQ3: Effects of the searchbot condition on participants' perceptions of the collaborative experience. Symbols `*', `**', and `***' denote significant differences at \emph{p} < .05, \emph{p} < .01, and \emph{p} < .001 level.}\label{fig:collab_exp}
\end{figure}

In RQ3, we investigate the effects of the searchbot condition on participants' post-task perceptions of their collaborative experience: (1) awareness of each other's activities, (2) ease of collaboration, and (3) enjoyment. Figure~\ref{fig:collab_exp} shows the means and 95\% confidence intervals across searchbot conditions.

\textbf{Collaborative Awareness:} The searchbot condition had a significant effect for two awareness measures: (1) awareness about my partner (${\chi(2)}^2$=15.411 , $p$<.001) and (2) my partner's awareness about me (${\chi(2)}^2$=15.439, $p$<.001). Participants reported being less aware of their partner's activities in the \textsc{BotTask} versus \textsc{BotInfo} ($\beta$=-0.630, S.E.=0.160, $p$<.001) and \textsc{BotDialog} condition ($\beta$=-0.463, S.E.=0.160, $p$<.01). Similarly, participants perceived their partners to be less aware of their own activities in the \textsc{BotTask} versus \textsc{BotInfo} ($\beta$=-0.722, S.E.=0.193, $p$<.001) and \textsc{BotDialog} condition ($\beta$=-0.611, S.E.=0.193, $p$<.01). Additionally, the searchbot condition had a marginally significant effect for joint attention---the extent to which participants looked at and discussed the \emph{same} information (${\chi(2)}^2$=5.31 , $p$ =.07). Participants found it harder to maintain joint attention in the \textsc{BotTask} versus \textsc{BotInfo} ($\beta$=-0.352, S.E.=0.174, $p$<.05) and \textsc{BotDialog} condition ($\beta$=-0.320, S.E.=0.174, $p$<.05).

\textbf{Collaborative Effort:} The searchbot condition had a significant effect for two measures associated with collaborative effort: (1) ease of sharing information (${\chi(2)}^2$=8.91, $p$<.05) and (2) having a smooth ``flow'' of communication (${\chi(2)}^2$=7.533, $p$<.05). Participants reported greater difficultly sharing information in the \textsc{BotTask} versus \textsc{BotInfo} ($\beta$=-0.444, S.E.=0.177, $p$<.05) and \textsc{BotDialog} condition ($\beta$=-0.489, S.E.=0.178, $p$<.01). Similarly, participants reported greater difficulty having a smooth ``flow'' of communication in the \textsc{BotTask} versus \textsc{BotInfo} ($\beta$=-0.537, S.E.=0.209, $p$<.05) and \textsc{BotDialog} condition ($\beta$=-0.469, S.E.=0.210, $p$<.05).

\textbf{Collaborative Enjoyment:} The searchbot condition had a significant effect on participants' own enjoyment (${\chi(2)}^2$=6.586, $p$<.05) and perceptions of their partner's enjoyment (${\chi(2)}^2$=6.062, $p$<.05). Participants reported enjoying themselves significantly more in the \textsc{BotInfo} versus \textsc{BotTask} condition ($\beta$=0.52, S.E.=0.210, $p$<.05). Similarly, participants perceived their partners to enjoy themselves more in the \textsc{BotInfo} versus \textsc{BotTask} condition ($\beta$=0.537, S.E.=0.218, $p$<.05).

\subsection{RQ4: Communication and Collaboration Measures}

In RQ4, we investigate the effects of the searchbot condition on measures related to participants' communication and collaboration. In terms of communication patterns between participants and the searchbot, we specifically focused on four measures (Section~\ref{sec:interactions}): (1) TME: total messages exchanged, (2) TLS: total links shared by the searchbot, (3) TEI: total number of times the participants explicitly interacted with the searchbot, and (4) TSI: total number of times the searchbot took the initiative.

\begin{figure*}[t]
	\centering
	\hspace{-1cm}
	\begin{subfigure}[c]{0.38\textwidth}
		\centering
		\includegraphics[width =\columnwidth]{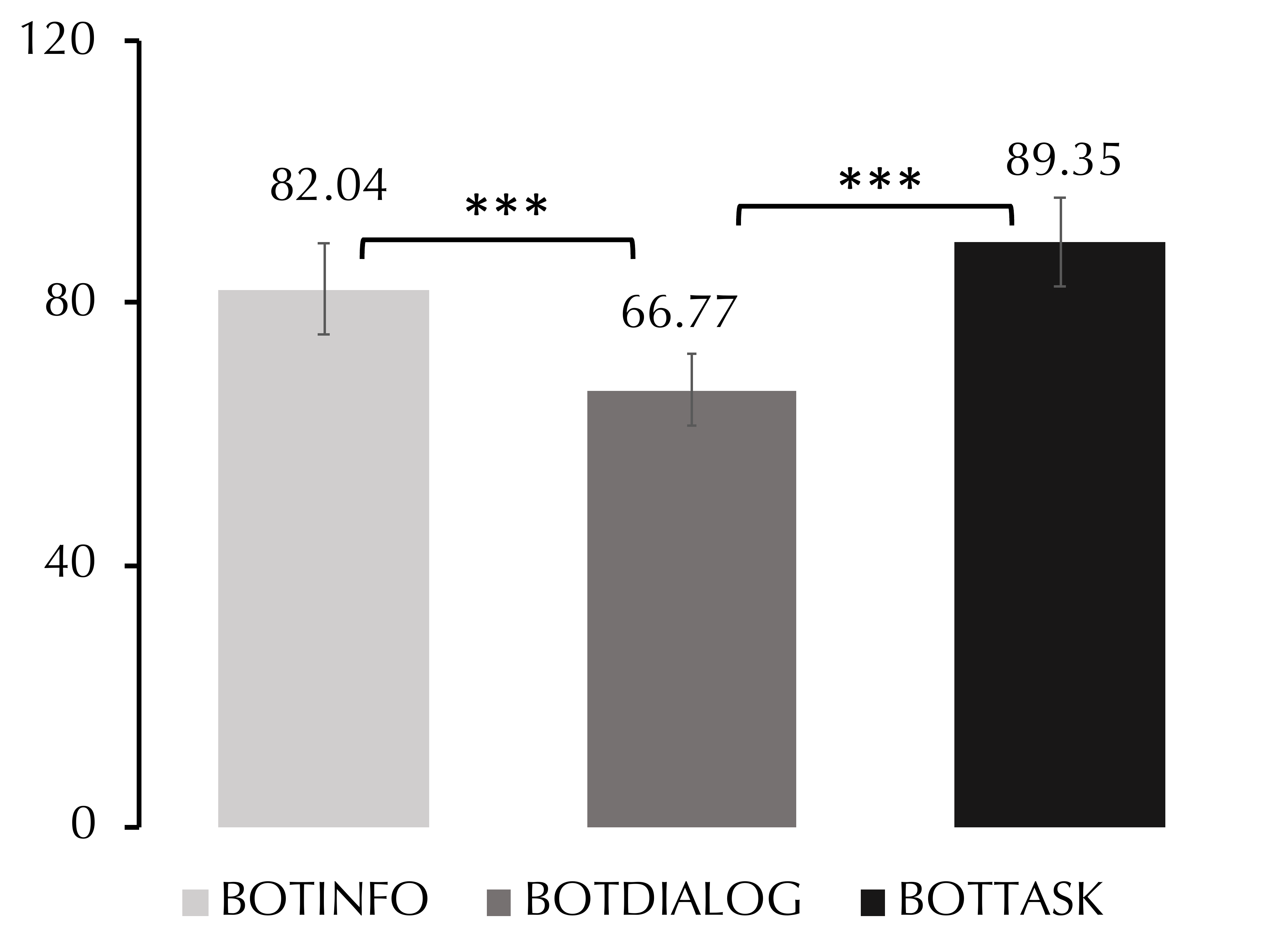}
		\caption{Total messages exchanged (TME).}
		\label{subfig:bot_info}
	\end{subfigure}
	\begin{subfigure}[c]{0.65\textwidth}
		\centering
		\includegraphics[width =\columnwidth]{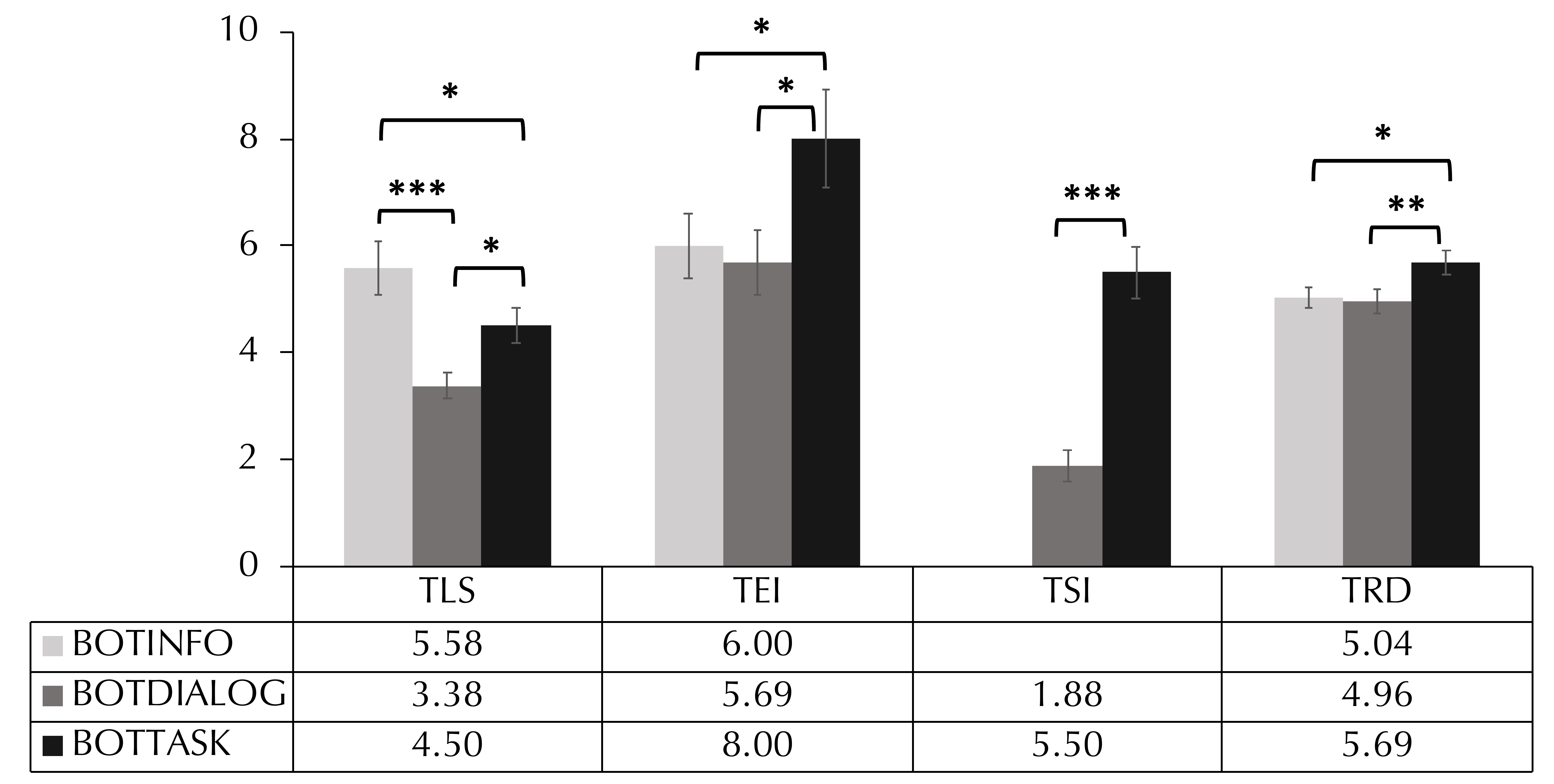}
		\caption{Total links sent (TLS), total explicit interactions (TEI), total searchbot initiatives (TSI), and total task-relevant dimensions considered (TRD).}
		\label{subfig:bot_dialog}
	\end{subfigure}	
	\caption{RQ4: Effects of the searchbot condition on participants' communication and collaboration patterns: total messages exchanged (TME), total web links sent by the searchbot (TLS), total explicit interactions with the searchbot (TEI), total number of times the searchbot took the initiative (TSI), and total number of task-relevant dimensions considered by participants (TRD). Symbols `*', `**' and `***' denote significant differences at \emph{p} < .05, \emph{p} < .01, and \emph{p} < .001 level.\vspace{-.3cm}}\label{fig:comm}
\end{figure*}

As shown in Figure~\ref{fig:comm}, the searchbot condition had a significant effect on all four measures: (1) TME (${\chi(2)}^2$=19.497, $p$<.001); (2) TLS (${\chi(2)}^2$=20.959, $p$<.01); (3) TEI (${\chi(2)}^2$=7.488, $p$<.05); and (4) TSI (${\chi(1)}^2$=42.764, $p$<.01). In terms of TME, participants exchanged significantly fewer messages in the \textsc{BotDialog} versus \textsc{BotInfo} ($\beta$=-15.154, S.E.=4.731, $p$<.01) and \textsc{BotTask} condition ($\beta$=-22.570, S.E.=4.731, $p$<.001). In terms of TLS, the searchbot shared significantly more results in the \textsc{BotInfo} versus \textsc{BotDialog} ($\beta$=2.159, S.E.=0.440, $p$<.001) and \textsc{BotTask} condition ($\beta$=1.113, S.E.=0.440, $p$<.05). Additionally, the searchbot shared significantly more results in the \textsc{BotTask} versus \textsc{BotDialog} condition ($\beta$=1.046, S.E.=0.444, $p$<.05). In terms of TEI, participants interacted more with the searchbot in the \textsc{BotTask} versus \textsc{BotInfo} ($\beta$=1.1918, S.E.=0.848, $p$<.05) and \textsc{BotDialog} condition ($\beta$=2.2148, S.E.=0.848, $p$<.01). Finally, in terms of TSI, the searchbot took significantly more initiative in the \textsc{BotTask} versus \textsc{BotDialog} condition ($\beta$=3.605, S.E.=0.443, $p$<.001).

\pagebreak
In terms of the collaboration, the searchbot condition had a significant effect on the number of task-relevant dimensions (TRD) considered by participants during the task (${\chi(1)}^2$=9.934, $p$<.01). As shown in Figure~\ref{fig:comm}, participants considered more dimensions in the \textsc{BotTask} versus \textsc{BotInfo} ($\beta$=0.669, S.E.=0.243, $p$<.05) and \textsc{BotDialog} condition ($\beta$=0.708, S.E.=0.243, $p$<.01).

\vspace{-.2cm}
\subsection{RQ5: Perceived Benefits}\label{subsec:rq5}
To investigate RQ5, we analyzed participants' responses to an open-ended question about perceived benefits from the searchbot (i.e., Max). We identified 10 codes associated with perceived benefits. Some codes were found in all three searchbot conditions. Other codes were found in some conditions and not others. We indicate the searchbot condition(s) associated with each code in brackets and the frequency of the code per condition in parentheses.

\textbf{Relevant Info [Info (31) + Dialog (35) + Task (37)]:} Participants mentioned that Max provided useful information during the task. For example, P5 said \emph{``[Max] found very helpful sites for us to go on and we were able to find everything we needed.''}

\textbf{Efficient [Info (10) + Dialog (11) + Task (10)]:} Participants described how Max helped make the search process more efficient. Participants felt that Max was efficient when she shared highly relevant information in a timely manner. For example, P25 said \emph{``[Max] was quick and provided answers that directly answered my questions.''} Participants also found Max to be efficient because she provided only one search result at a time. For example, P4 said \emph{``[Max] limited my options by only showing me one result so I wouldn't spend a long time looking.''}

\textbf{Ideas [Info (1) + Dialog (4) + Task (11)]:} Participants described how Max helped them broaden their perspectives on the task. Participants reported that Max provided alternative ideas regarding options and dimensions relevant to the task. In the \textsc{BotInfo} condition, this was done in a relatively \emph{indirect} manner, by sharing search results that included alternative ideas. In the \textsc{BotDialog} condition, Max sometimes asked clarification questions that introduced new ideas. For example, P51 said \emph{``Max asked what kinds of fillings we needed when we did not even think about that.''} In the \textsc{BotTask} condition, Max could also provide new ideas by explicitly making suggestions. For example, P18 said \emph{``[Max] helped guide our search by pointing out allergens that we didn't consider initially in our first choice of recipe.''} Another participant (P3) even mentioned Max providing \emph{``countersuggestions''}. 

\textbf{Appreciation [Info (2) + Dialog (1) + Task (2)]:} Participants reported appreciating Max's efforts. Regardless of how much Max helped participants during the task, participants recognized that she tried to provide the best assistance possible. For example, P37 said \emph{``[Max] did their best to find the information that was needed, though it seemed limited.''}

\textbf{Facilitate Collaboration [Info (1) + Dialog (2) + Task (2)]:} Participants reported that Max helped facilitate the collaboration by: (1) keeping the conversation going and (2) helping with the decision-making process. In the \textsc{BotInfo} condition, providing search results in a timely manner appeared to be the determinant factor. For example, P41 said \emph{``[Max] gave us quick responses to our questions, which helped the flow of the conversation.''} In the \textsc{BotDialog} and \textsc{BotTask} conditions, participants reported that clarification questions helped with decision making. For example, P43 said \emph{``They provided good resources and asked for clarification that ultimately benefited our collaboration and decision making.''}

\textbf{Narrow/Refine [Dialog (9) + Task (9)]:} Participants mentioned that Max helped narrow/refine their searches. In both the \textsc{BotDialog} and \textsc{BotTask} conditions, participants reported that clarification questions helped narrow their focus. For example, in the \textsc{BotDialog} condition, P37 said \emph{``They provided relevant information and asked clarifying questions that helped to refine our search, which made it easier to find the information we wanted.''} In the \textsc{BotTask} condition, this could also happen when Max provided suggestions. For example, P25 said \emph{``When Max gave her opinions, it helped us fine-tune what we were looking for.''}

\textbf{Proactivity [Task (3)]:} Participants liked that Max played the role of a proactive collaborator, especially when they felt stalled during the task. For example, P50 said \emph{``I enjoyed that Max could join the conversation and narrow our [searches].''} Another participant (P16) said \emph{``[Max] offered suggestions when we were stuck and was helpful in identifying different programs.''} 

\textbf{Check-in [Task (1)]:} One participant mentioned that Max checked in with them when they were silent. For example, P4 said \emph{``Max followed up if we did not respond back to her after a minute.''}

\textbf{Auxiliary Info [Task (3)]:} Participants mentioned that Max provided \emph{additional} information that was not requested but useful. For example, P3 said \emph{``[Max] brought up relevant surrounding information and suggestions''} Another participant (P40) said \emph{``[Max] was able to give us additional information that helped us make decisions.''}

\textbf{Incorporated Feedback [Task (2)]:} Participants liked that Max incorporated their feedback throughout the task. For example, P6 said \emph{``[Max] provided useful information and listened to our input.''}

\subsection{RQ6: Perceived Challenges}\label{subsec:rq6}

To investigate RQ6, we analyzed participants' responses to an open-ended question about perceived challenges while interacting with the searchbot (i.e., Max). We identified 12 codes associated with perceived challenges.

\textbf{Poor Results [Info (9) + Dialog (6) + Task (3)]:} Participants mentioned that Max shared information that was not ``on point''. Specifically, participants found a shared search result not helpful when: (1) it was not specific enough or (2) it did not incorporate previously mentioned needs. For example, P4 said \emph{``The question was slightly misinterpreted and the response was too broad.''} Another participant (P2) said \emph{``I specified food allergies after Max asked for clarification but the link she sent was about regular [not food-related] allergies.''}

\textbf{Partial Coverage [Info (3) + Dialog (3) + Task (1)]:} Participants mentioned that Max provided information that only \emph{partially} addressed their needs or was somehow limited. This occurred when Max provided information that failed to meet all the desired criteria. For example, P29 said \emph{``Max provided options that didn't fulfill all of our criteria or it didn't have all the information we needed. It offered places that were good and cheap, but not places that were safe.''} Additionally, participants mentioned that they expected more relevant information to exist on the web. For example, P37 said \emph{``I feel as if there may have been more relevant information to our questions, but we did not receive it.''}

\textbf{Lack of Direct Control [Info (2) + Dialog (3) + Task (1)]:} Participants mentioned that relying on Max made them feel a lack of agency. For example, P1 said \emph{``It would've been much faster to search on our own and also have many search results to choose the best.''} Another participant (P11) said \emph{``I feel that I would've not been able to leave the situation satisfied had I relied fully on the bot. The only reason I felt comfortable with our decision was because my partner had pre-existing knowledge and used the bot for targeted searches.''}

\textbf{Made it Challenging [Info (2) + Dialog (2) + Task (2)]:} Participants mentioned that Max made the task more difficult. Participants reported expending effort on: (1) accurately communicating their needs to Max and (2) processing the information shared by Max when it was not presented the way they wanted. For example, P21 said \emph{``Max could not find exactly what I needed despite trying a couple of times.''} Another participant (P6) said \emph{``I really wanted a map that had the cities on it, which was not what was pulled up.''} 

\textbf{Delay in Response [Info (4) + Dialog (3) + Task (3)]:} Participants complained about Max taking too much time to return results. For example, P50 said \emph{``Max took too long to find what I needed.''}

\textbf{Inconsistent Performance [Info (1) + Task (1)]:} Participants mentioned Max's performance being inconsistent throughout the task. For example, P45 said \emph{``[Max] felt like Siri. Siri sometimes hits the spot with the answers, but sometimes she doesn't.''} 

\textbf{Excessive Follow-up [Dialog (2) + Task (2)]:} Participants complained about the time delay caused by Max's follow-up responses after a request (i.e., clarification questions or suggestions). For example, P2 said \emph{``Much time was spent by [Max] trying to understand the question instead of providing information.''} Another participant (P1) said \emph{``Max took a lot of time sending a helpful link because they asked too many questions.''}

\textbf{Lack of Technical Knowhow [Dialog (1)]:} One participant complained about not knowing how to modify or update an existing request. For example, P31 said \emph{``When we wanted to change the question we asked, we were unsure how to do that and we spent time deliberating what to do about the found answer to our question.''} This indicates that searchers may want to perform complex operations (e.g., cancellations or modifications) on information requests that are already being processed by the searchbot.

\textbf{Missed Intervention [Dialog (1)]:} One participant pointed out that Max missed an opportunity to intervene and be helpful. For example, P24 said \emph{``There was one question we asked that was pretty vague that could have benefited from clarification questions.''}

\textbf{Stale Intervention [Task (1)]:} One participant mentioned that Max made suggestions that were no longer relevant at the time of the intervention. For example, P40 said \emph{``Occasionally when I felt as though we had moved on from a topic of discussion, Max would still give suggestions based on that topic, which was a bit of a distraction.''}

\textbf{Disruption [Task (3)]:} Participants mentioned Max's interventions being disruptive. For example, P20 said \emph{``Max popped in at random times that we did not need help.''} Interestingly, it was not only the inappropriate timing or unclear intent of the intervention that made participants feel disrupted. Participants complained about the mere fact that Max made suggestions \emph{without being asked}. Participants also reported that proactive suggestions made them think that Max lacked confidence in their ability. For example, P1 said \emph{``[Max] interrupted with suggestions without being asked [...] and was not confident in our decision making.''}

\textbf{Feeling Spied On [Task (2)]:} Participants expressed discomfort with Max monitoring their conversation. For example, P20 said \emph{``It felt like they were spying on us too much.''} Another participant (P21) said \emph{``At times Max felt intrusive [...] listening in on our conversation.''}
\vspace{-.3cm}
\section{Discussion}\label{sec:discussion}
\vspace{-.1cm}
Taken together, our RQ1-RQ6 results highlight potential benefits and challenges for searchbots that can take different levels of initiative to support users during collaborative search tasks.  In this section, we summarize the main trends observed in our results and relate these trends to prior research.  RQ1-RQ3 focused on participants' perceptions, RQ4 focused on participants' behaviors, and RQ5-RQ6 focused on participants' responses to open-ended questions about the benefits and challenges of interacting with the searchbot in a specific condition.  We begin this section by discussing trends observed in our RQ1-RQ3 results.  To help explain our RQ1-RQ3 results, we make connections with results observed for RQ4-RQ6 and also draw on insights gained by analyzing the conversations captured within the Slack channel.  We end this section by discussing important benefits and challenges reported by participants in different searchbot conditions. 

\textbf{Coverage (RQ1, RQ4, RQ5):} \hl{Based on our RQ1 results, participants perceived that they covered less task-relevant information when the searchbot could make task-level suggestions (\textsc{BotTask}) versus only ask clarification questions in response to a request (\textsc{BotDialog}).} This effect on participants' \emph{perceptions} may seem counterintuitive when we also consider our RQ4 and RQ5 results. In terms of RQ4, in the \textsc{BotTask} condition, participants considered more task-relevant dimensions that were not included in the task description (Figure~\ref{fig:comm}). Similarly, in terms of RQ5, more participants commented that the searchbot provided more ``new ideas'' for them to consider during the task.

An important follow-up question is: Why did participants perceive to cover \emph{less} task-relevant information when they considered \emph{more} task-relevant dimensions and \emph{more} new ideas during the \textsc{BotTask} condition? One possible explanation is that perceived coverage is a function of the \emph{information space covered compared to the total information space}. In this respect, task-level suggestions may have led participants to realize that the total information space of the task was \emph{larger than they originally conceived}. This realization may have led to reduced perceptions of coverage.

\begin{figure}[t]
	\centering
	\includegraphics[width=0.9\textwidth]{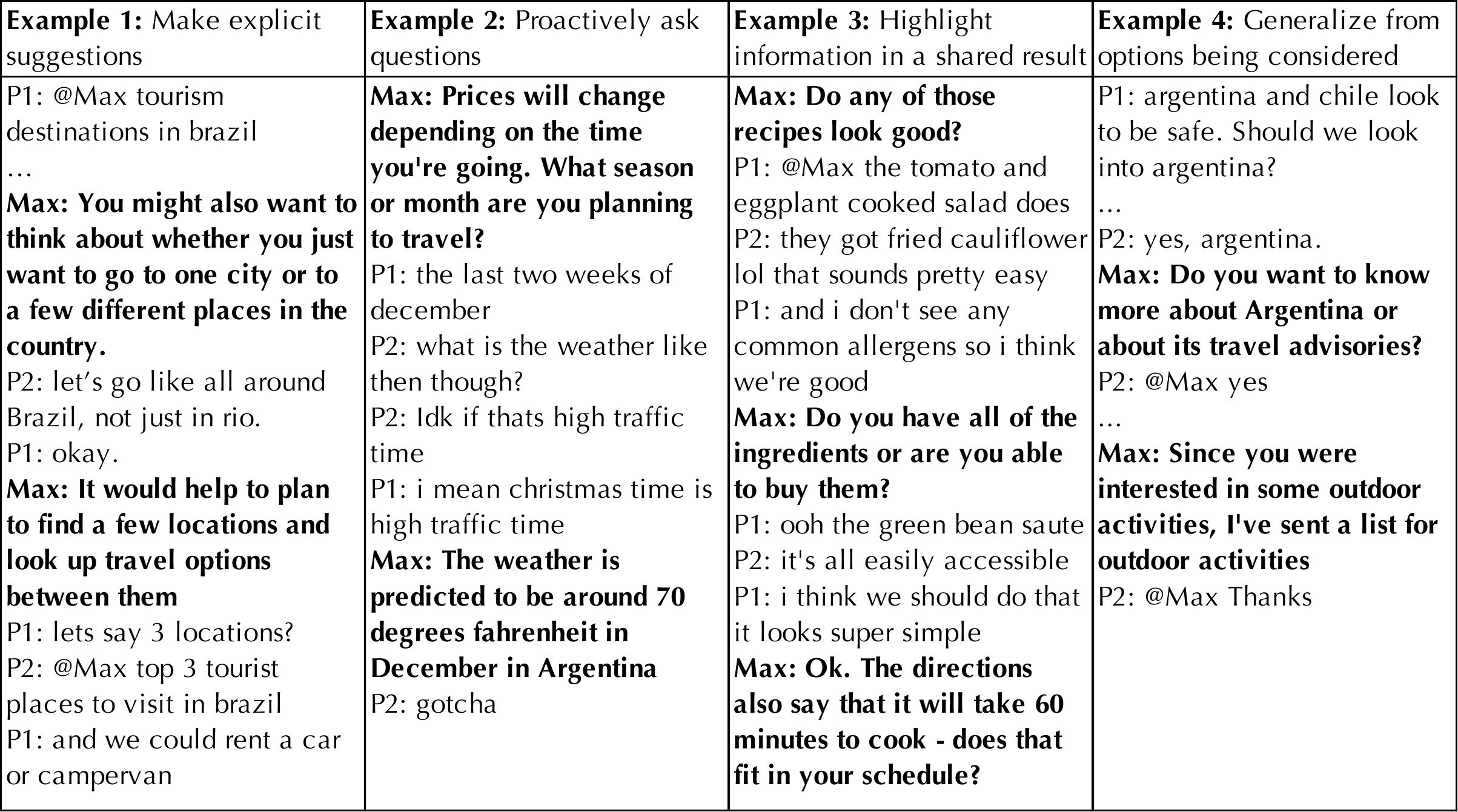}
	\vspace{-.2cm}
	\caption{Examples of the searchbot in the \textsc{BotTask} condition influencing participants to consider new task-relevant dimensions.\vspace{-.5cm}}\label{fig:example}
\end{figure}

\textbf{Task-Relevant Dimensions (RQ4):} As previously mentioned, in the \textsc{BotTask} condition, participants considered more task-relevant dimensions that were not included in the task description (Figure~\ref{fig:comm}). We wanted to better understand the \emph{mechanisms} through which this was achieved. By analyzing the chatlogs captured over Slack, we observed four different mechanisms through which the searchbot led participants to consider more task-relevant dimensions in the \textsc{BotTask} condition. Figure~\ref{fig:example} illustrates one example of each case (Example 1-4). 

First, the searchbot sometimes \emph{explicitly} recommended that participants consider a new dimension they had not considered. In Example 1, the searchbot explicitly asked participants to consider travel destinations where they could visit different nearby cities or landmarks. Here, the searchbot led participants to consider the ``proximity of attractions'' dimension. Second, in some cases, the searchbot asked questions that prompted participants to consider a new dimension. Such questions were primarily aimed at better understanding the participants' needs and preferences. However, as an unexpected side effect, they also led participants to discuss a new dimension. In Example 2, the searchbot asked participants about when they planned to travel. Ultimately, this question led participants to discuss the ``weather'' dimension. Third, after sharing a link, the searchbot sometimes followed-up to highlight an important (new) dimension associated with the information sent. In Example 3, after sharing a link, the searchbot asked participants whether they had access to all the ingredients in the recipe and whether they had 60 minutes available to make the dish. This information prompted participants to consider two new dimensions: ``availability of ingredients'' and ``cooking time''. Finally, in some cases, the searchbot highlighted a common characteristic of options being considered by participants. In Example 4, the searchbot noticed that participants were considering travel destinations with ``outdoor activities'' and highlighted this dimension as a rationale to recommend other destinations.

\textbf{Disruption (RQ1, RQ6):} \hl{Participants perceived the searchbot to be \emph{more disruptive} during the \textsc{BotTask} versus \textsc{BotInfo} and \textsc{BotDialog} condition (Figure~\mbox{\ref{fig:utility}}).} An important follow-up question is: Why was the searchbot perceived to be more disruptive in the \textsc{BotTask} versus \textsc{BotDialog} condition? There are five possible reasons for why task-level suggestions (at times provided proactively) were more disruptive than clarification questions.

First, the searchbot typically asked clarification questions shortly after participants issued an information request. Thus, clarification questions were typically asked when \emph{both} participants were engaged with the searchbot versus engaged in other activities (e.g., individually reading web pages or jointly discussing options over Slack). In other words, clarification questions were typically asked when the attention of \emph{both} participants was focused on the searchbot (i.e., as participants waited for the searchbot to return information).

Second, prior work has found that system interventions are more disruptive when users are cognitively engaged in a task and less disruptive when users are transitioning between subtasks (i.e., points of low cognitive load)~\cite{Adamczyk2004,Iqbal2006,Iqbal2008}. It is possible that participants sent information requests to the searchbot at points when they were transitioning from one subtask to the next. In this respect, clarification questions may have been asked when both participants had the cognitive resources available to address them promptly and effectively.

Third, by definition, clarification questions have a clear objective---to elicit additional information for the searchbot to better understand the information need behind a request. It is possible that some of the task-level suggestions made by the searchbot in the \textsc{BotTask} condition did not reveal a clear objective, readily understood and valued by the participants. Amershi et al.~\cite{amershi2019guidelines} argued that mixed-initiative systems should always allow users to access the \emph{rationale} underlying a specific system intervention. We believe that some of the task-level suggestions provided by the searchbot did \emph{not} include a clear explanation of the searchbot's rationale for making the suggestion.

Fourth, it is possible that the searchbot made task-level suggestions that were not relevant and therefore perceived as disruptive. Indeed, in our RQ6 results, one participant mentioned that the searchbot made suggestions on topics that were no longer relevant. In this case, the Wizard possibly missed a topic transition when monitoring the participants' conversation over Slack.

Finally, in the \textsc{BotTask} condition, the searchbot could proactively make suggestions at any point without being asked. Task-level suggestions were intended to influence the participants' goals and/or approach to the task. It is possible that some proactive suggestions were made too early or too late during the session. Avula et al.~\cite{avula2018searchbots} investigated searchbots that can proactively intervene to resolve an implicit information need detected in the conversation. Participants perceived these interventions to be disruptive when they occurred too soon (i.e., before the participants fully understood the task requirements) or too late (i.e., after participants had already committed to a specific approach to the task). In this respect, task-level suggestions may need to occur after collaborators understand the task requirements and before they commit to a specific approach.

\textbf{Effort (RQ2, RQ4-RQ6):} Based on our RQ2 results, participants reported less effort in the \textsc{BotDialog} versus \textsc{BotInfo} and \textsc{BotTask} condition (Figure~\ref{fig:workload}). Our RQ4-RQ6 results suggest that the \textsc{BotInfo} and \textsc{BotTask} conditions required more effort for different reasons.

\hl{There are two possible reasons for why the \textsc{BotInfo} condition required more effort than the \textsc{BotDialog} condition. Based on our RQ5 results, participants reported that clarification questions helped narrow/refine their searches. In the \textsc{BotInfo} condition, participants had to narrow/refine their searches by reformulating their requests rather than responding to clarification questions. It is possible that answering clarification questions required less effort than reformulating requests without any feedback about \emph{why} or \emph{how} the original request was either too broad, specific, vague, or ambiguous.}

Second, our results suggest that clarification questions enabled the searchbot to find more relevant results. Based on our RQ5 results, more participants mentioned that the searchbot found ``relevant information'' during conditions where it took more initiative (\textsc{BotInfo} < \textsc{BotDialog} < \textsc{BotTask}). Similarly, based on our RQ6 results, more participants complained that the searchbot provided ``poor results'' in conditions where it took less initiative (\textsc{BotInfo} > \textsc{BotDialog} > \textsc{BotTask}). Thus, participants in the \textsc{BotInfo} condition may have expended more effort sifting through non-relevant results provided by the searchbot.

There are four possible reasons for why the \textsc{BotTask} condition required more effort than the \textsc{BotDialog} condition. First, as discussed above, participants perceived the searchbot to be the most disruptive in the \textsc{BotTask} condition. Above, we argued that proactive suggestions were likely to be disruptive when they occurred: (1) too soon, (2) too late, and (3) at times when participants focused on other tasks. Thus, it is likely that participants in the \textsc{BotTask} condition had to expend extra effort disengaging and reengaging with activities after being disrupted by untimely task-level suggestions. 

\hl{Second, based on our RQ4 and RQ5 results, participants considered more task-relevant dimensions and gained more new ideas in the \textsc{BotTask} condition. While new ideas can be beneficial, they may have also led to more deliberations between participants, resulting in greater effort.}

\hl{Third, in the \textsc{BotTask} condition, the searchbot was a more active participant in the collaboration. Based on our RQ4 results (Figure~\mbox{\ref{fig:comm}}), the \textsc{BotTask} condition was associated with the greatest amount of communication over Slack. It is likely that this greater level of communication resulted in greater perceptions of effort.} 

Finally, in the \textsc{BotTask} condition, the searchbot made task-level suggestions that were helpful but required participants to expend more effort. We observed this happening in four ways. First, the searchbot sometimes \emph{expanded} the set of options for participants to consider. Second, the searchbot sometimes recommended that participants \emph{re-direct} their approach to the task. Third, the searchbot sometimes reminded participants of preferences they had already discussed. For example, the searchbot sometimes pointed out when participants were considering options that were \emph{inconsistent} with criteria previously mentioned as being important. Finally, some task-level suggestions aimed to raise awareness about things participants might have missed (e.g., ``Did you notice that this recipe involves lots of ingredients?''). In all four cases, the task-level suggestions were intended to be helpful, but required participants to expend more effort. In other words, these task-level suggestions discouraged participants from \emph{satisficing}.

\textbf{Frustration \& Enjoyment (RQ2, RQ3, RQ6):} Based on our RQ2 and RQ3 results, participants reported greater frustration and lower enjoyment in the \textsc{BotTask} condition (Figures~\ref{fig:workload} \&~\ref{fig:collab_exp}). There are several possible reasons for this trend. First, some of the frustration experienced by participants in the \textsc{BotTask} condition can be explained by previously discussed results. For example, participants in the \textsc{BotTask} condition reported expending more effort but \emph{also} perceived covering less task-relevant information. 

Second, in the \textsc{BotTask} condition, the searchbot had the greatest flexibility in how to assist participants. This may have caused participants to have higher (or even unreasonable) expectations about the searchbot's capabilities that were difficult to meet. Indeed, prior research has found that raising a user's expectations about a system's competence can backfire and lead to lower perceptions of usability~\cite{Khadpe2020}. 

\hl{Finally, based on our RQ6 results, a few comments by participants point to more nuanced sources of frustration such as the feeling of being ``spied on'' or the searchbot's ``lack of confidence'' in participants' ability to make good decisions. This result suggests that some participants may prefer to be in control of their approach to the task, even if it is suboptimal. Indeed, prior research has found that users often avoid dynamic help systems simply because they ``refuse to admit defeat''~\mbox{\cite{Dworman2004}}.}

\textbf{Collaborative Awareness (RQ3, RQ4, RQ5):} Based on our RQ3 results, participants reported less awareness of each other's activities in the \textsc{BotTask} versus \textsc{BotInfo} and \textsc{BotDialog} condition (Figure~\ref{fig:collab_exp}). At first, this trend may seem to contradict results from previous studies. Avula et al.~\cite{avula2018searchbots,Avula2019} investigated the impact of different searchbot designs on participants' collaborative awareness, and found that allowing participants to search directly from the Slack channel improved participants' awareness of each other's activities. However, these studies considered searchbots that can only ask clarification questions and not proactively intervene to provide task-level advice.

There are three possible reasons for why participants may have experienced lower collaborative awareness in the \textsc{BotTask} condition. \hl{First, our RQ4 results found that there were more messages exchanged between the participants and the searchbot in the \textsc{BotTask} condition. Thus, participants might have devoted more attention to the searchbot, which may have reduced their attention to each other.}

\hl{Second, the searchbot in the \textsc{BotTask} condition might have provided more ``food for thought'' by introducing new ideas. Participants in the \textsc{BotTask} condition had to consider the relevance of these new ideas \emph{individually} before potentially incorporating them into their plans. Thus, participants may have paid less attention to each other's activities while individually considering these ideas.}

Finally, in a sense, the searchbot during the \textsc{BotTask} condition played the role of a \emph{third} collaborator. Prior research in group dynamics has found that increasing the group size can lead to lower mutual awareness, and that lower mutual awareness can in turn lead to lower trust and commitment to the group~\cite{Soboroff2020}.

\textbf{Benefits of Dialog-level Initiative (RQ5, RQ6):} \hl{Our RQ5 and RQ6 results suggest that dialog-level initiative provided important benefits. As expected, asking clarification questions helped participants ``narrow/refine'' their searches (RQ5) and enabled the searchbot to find more relevant information (RQ6).}

Interestingly, our RQ5 results also suggest that clarification questions provided \emph{indirect} benefits: (1) provided new ideas, (2) stimulated the conversation, and (3) helped participants with the decision-making process. Some clarification questions elicited additional \emph{necessary} information (e.g., ``Visa requirements vary by nationality. What country are you from?''). However, other clarification questions asked participants about: (1) specific preferences (e.g., ``What \underline{type of volunteering opportunity} are you interested in?''); (2) specific constraints (e.g., ``What is your \underline{budget}?''); and (3) dimensions of the task they had not explicitly considered (e.g., ``Do you want recipes with ingredients that are \underline{easy to find}?''). Our results suggest that such clarification questions helped collaborators become aware of things they needed to consider and helped stimulate the discussion.

\textbf{Benefits of Task-level Initiative (RQ4, RQ5):} \hl{Our RQ4 and RQ5 results suggest that task-level suggestions provided four important benefits. First, in the \textsc{BotTask} condition, the searchbot provided more ``new ideas'' for participants to consider while comparing and evaluating options.} 

Second, participants mentioned that they appreciated the searchbot \emph{proactively} intervening when they ``were stuck''. In the context of human-robot interaction, Jiang and Arkin~\cite{jiang2015mixed} argued that proactive system interventions can be specifically designed to help with three general processes: (1) developing objectives, (2) developing strategies to accomplish an objective, and (3) executing strategies to meet an objective. Clarification questions typically help with the third process: executing a strategy. Our results suggest that task-level suggestions can help with the first two processes: goal development and strategy planning. This is an important benefit of task-level suggestions. More than four decades ago, Carbonell~\cite{carbonell1970ai} argued that mixed-initiative AI systems are well-positioned to help users \emph{diagnose} problems and \emph{plan} a solution.

\hl{Third, in the \textsc{BotTask} condition, the searchbot sometimes provided ``auxiliary information'' associated with a link shared in response to a request. We observed two mechanisms through which this was achieved. In one case, the searchbot provided relevant information that was not explicitly requested. For example, when participants were considering traveling to Peru, the searchbot interjected to notify participants about protests happening in Peru during that time. In the second case, the searchbot interjected to correct important misconceptions. For example, when participants seemed to think that Nairobi and Kenya were different destinations, the searchbot interjected to clarify that Nairobi is the capital of Kenya.}

Finally, in the \textsc{BotTask} condition, participants commented that they appreciated the searchbot incorporating their feedback \emph{throughout the task}. Prior work also found that users value when a conversational agent incorporates their feedback throughout the entire search session~\cite{vtyurina2017exploring}.

\textbf{Challenges from Dialog-level Initiative (RQ6):} Our RQ6 results found three important challenges for systems that ask clarification questions. First, two participants complained about the searchbot asking too many clarification questions. This result resonates with previous recommendations that interventions should be limited to cases where the intervention has \emph{obvious} added value~\cite{horvitz1999principles}. Second, one participant complained about the searchbot missing an opportunity to ask a clarification question in response to a vague request. This result resonates with previous recommendations that systems should intervene when the benefits outweigh the costs~\cite{horvitz1999principles,amershi2019guidelines}. Finally, one participant complained about not knowing how to \emph{modify} a request that was already being processed by the searchbot. To avoid such scenarios, prior work has recommended that systems should adequately convey what they can and cannot do~\cite{amershi2019guidelines}. It is interesting that this complaint was only present in the \textsc{BotDialog} condition. It may be especially important to convey a system's capabilities in cases where the mixed-initiative system is \emph{limited} in the types of initiative it can take (i.e., when the system can do some things but not others).

\textbf{Challenges from Task-level Initiative (RQ6):} Our RQ6 results found two important challenges for systems that can proactively give task-level advice. It is noteworthy that systems are likely to face both challenges even when they provide task-level suggestions that are relevant and timely.

First, in the \textsc{BotTask} condition, two participants complained about feeling ``spied on''. These participants felt uncomfortable with the searchbot ``listening in'' on their conversation. Systems that proactively intervene with task-level advice will need to monitor the conversation to some extent. To address such privacy concerns, we see two possible paths forward. Perhaps the easiest solution is to allow users to disable task-level suggestions on certain collaborations. Amershi et al.~\cite{amershi2019guidelines} proposed that mixed-initiative systems should always allow users to change privacy permissions and allow ``private mode''. A second solution is to limit task-level suggestions to those less likely to be perceived as ``creepy''. Recent research has investigated factors that contribute to the ``creepiness factor'' of personalized recommendations~\cite{Torkamaan2019}. Results suggest that personalized recommendations are more ``creepy'' in certain domains and in the presence of causal ambiguity. In this respect, task-level suggestions may be perceived as less ``creepy'' when the conversation is not on a sensitive subject and when the system can describe the evidence and rationale for making a suggestion.

Second, in the \textsc{BotTask} condition, several participants complained about the searchbot intervening ``without being asked'' or ``when not invited''. To address these concerns, we see two possible directions to explore. One alternative is to limit task-level suggestions to only those that are critical for task success, and to accompany each suggestion with a clear rationale for why it is important. Another alternative is to adjust the level of advice on a case-by-case basis. In the context of intelligent tutoring systems, Graesser et al.~\cite{graesser2005autotutor} proposed that system interventions are more likely to be effective in domains where the learner has low prior knowledge. In this respect, collaborators may be more receptive to task-level suggestions when they have low domain knowledge.

\section{Implications for Future Work}\label{sec:implications}

Designing mixed-initiative conversational agents to support collaborative information seeking is a complex challenge. Our study identified different benefits and challenges of a system that can take initiative at different levels (i.e., dialog- and task-level initiative). Based on our findings, we discuss opportunities and implications for future work. 

\textbf{Dialog-level Initiative:} Our results found that asking clarification questions has four main benefits. First, as expected, they can help collaborators refine their searches and help the system find more relevant results. Second, while clarification questions are primarily aimed at better understanding the current need, they can also provide collaborators with new ideas. Third, they can reduce the level of effort required by the task. Finally, they can be less disruptive than proactive suggestions. This is likely due to the timing of clarification questions and their inherent nature. In terms of their timing, clarification questions are asked shortly after an information request. In this respect, they are likely asked when collaborators are engaged with the searchbot (vs. other activities) and at times when collaborators are transitioning between subtasks (i.e., points of low cognitive load). In other words, clarification questions are likely asked when participants will notice them and have the cognitive resources to address them. In terms of their nature, clarification questions have a clear \emph{implicit} objective (i.e., to better understand a request) and are directly aligned with the collaborators' current goal (i.e., to find specific information). In other words, compared to task-level suggestions, collaborators are likely to understand \emph{why} a clarification question is being asked and \emph{how} it might help them move forward with the task.

Our results found several challenges for dialog-level initiative that need to be addressed in future work. First, consistent with recommendations from prior work, systems should ask clarification questions by weighing their benefits and costs~\cite{horvitz1999principles,amershi2019guidelines}.  Importantly, to avoid asking too many questions, systems should consider previous interactions when estimating the cost of a new clarification question. Second, systems should be able to communicate to users how they can interact with the searchbot (i.e., what operations they can perform on it). Prior work has argued that mixed-initiative systems should clearly convey what they can (and cannot) do and how~\cite{amershi2019guidelines}.  We believe this is especially critical for mixed-initiative systems that are conversational (i.e., lack a visual interface to convey acceptable user actions) and are limited in their dialog capabilities (i.e., can respond to and perform certain actions but not others). Interestingly, in our study, participants mentioned deliberating on how to perform a specific action with the searchbot (e.g., how to modify a request). Thus, as an implication for future work, systems may be able to analyze conversations to gain insights about users' mental models and expectations of the system.

\textbf{Task-level Initiative:} In recent years, the IR research community has mostly focused on developing conversational search systems that can elicit additional information about a searcher's need---a form of dialog-level initiative. However, further into the future, we will likely see systems that can take task-level initiative to support either a single searcher or, as in our study, multiple collaborators. 

\hl{We found three main benefits of making task-level suggestions. First, task-level suggestions can provide collaborators with new ideas.  Second, task-level suggestions can help collaborators become ``unstuck'' when they do not know how to proceed with the task. This is a unique benefit of task-level suggestions. While clarification questions can help with strategy execution, task-level suggestions can also help with goal setting and strategy planning~\mbox{\cite{jiang2015mixed}}. Third, task-level suggestions can provide useful information that was not explicitly requested. In our study, we observed cases where the searchbot made proactive suggestions to: (1) provide useful background information about options being considered (e.g., ``You might want to re-consider option ABC because of XYZ.'') and (2) correct misconceptions as evidenced in the collaborators' conversation (e.g., ``Option ABC and XYZ are actually the same.'').}

Our results also found several challenges for systems that can take task-level initiative to support collaborative search. These insights are an important contribution of our work. First, while proactive suggestions can provide benefits, they can also be disruptive. To alleviate this drawback, future research will need to work on the timing, relevance, and delivery of proactive suggestions. In terms of their timing, proactive suggestions may be less disruptive when collaborators are engaged with the system (e.g., shortly after an information request), during times of low cognitive load (e.g., during subtask transitions), and when the suggestions are neither too early (e.g., before collaborators fully understand the task) nor too late (e.g., after collaborators commit to an approach).  In terms of their relevance, systems may need to carefully monitor the conversation and make suggestions that are relevant to the current objective (i.e., not on topics collaborators have already abandoned) and suggestions that are consistent with preferences/constraints mentioned in the conversation.  In terms of their delivery, proactive suggestions may \emph{not} be immediately understood and valued by collaborators. Thus, systems may need to accompany suggestions with a clear justification for why they are being made and how they might help.

Second, our results suggest that task-level suggestions can lead to greater effort. Greater effort is not necessarily a negative outcome. For example, task-level suggestions can help collaborators consider a wider range of options and criteria, as well as adopt better approaches to the task.  However, system designers should be mindful that task-level suggestions may incur a greater cost than clarification questions.

Third, our results suggest that task-level suggestions can lead to greater frustration and lower enjoyment.  This can be partly attributed to proactive suggestions being more disruptive (discussed above). Additionally, we believe that task-level suggestions may also have unintended effects that can lead to greater frustration and lower enjoyment. Specifically, task-level suggestions may cause participants to have unreasonable expectations about the system's capabilities, as well as unreasonable expectations about what it means to complete the task successfully. Future work must address these challenges by designing systems that can influence users to have reasonable expectations about the system and their own performance.

Fourth, our results suggest that task-level suggestions can negatively impact certain perceptions of the collaborative experience, including collaborative awareness. \hl{As expected, task-level suggestions provided participants with new ideas for them to consider and potentially incorporate into their approach to the task.  While new ideas can provide benefits, they also need to be considered individually by each collaborator and subsequently discussed amongst collaborators. During a collaboration, new ideas may lead to unresolved discussions or disagreements. To address this challenge, systems may need to estimate the difficulty of incorporating specific suggestions into action plans and estimate the extent to which specific suggestions are consistent with the preferences and constraints of \emph{all} collaborators.}

Finally, our results suggest that task-level suggestions can negatively impact collaborators' sense of \emph{privacy} and \emph{agency}.  These are challenges that are likely to be present even if the system makes proactive suggestions that are timely and relevant.  To address these challenges, future research should investigate factors that impact these perceptions.  In terms of privacy, task-level suggestions may be perceived as less ``creepy'' if the conversation is not on a sensitive subject and if suggestions are accompanied with an explanation of the evidence used to make the suggestion (i.e., reducing causal ambiguity).  In terms of agency, system designers may need to carefully consider how the system communicates with users.  Task-level suggestions may need to be communicated in a way that helps collaborators retain a sense of control and a sense of not being judged by their performance.


\section{Caveats \& Limitations}

\hl{Our study involved certain experimental decisions that may have influenced our results.}

\hl{First, across all three searchbot conditions, participants were fully aware they were interacting with a human searchbot (i.e., a reference librarian).  In fact, before each task, participants were described the searchbot's capabilities associated with the next condition in the presence of the Wizard so that they could ask any clarification questions.  In Section~\mbox{\ref{subsec:rationales}}, we explain why we made this decision.  This decision may have influenced some of our results.  For example, perhaps knowing that the searchbot was human led participants to have unreasonably high expectations about the searchbot's capabilities and their own performance during the task. This may have contributed to greater levels of frustration during the task. Similarly, knowing that the searchbot was a human may have contributed to participants having lower perceptions of privacy and agency.}

\hl{Second, participants were given 15 minutes to complete each task.  This time constraint was imposed to keep the entire study session under two hours.  Our results suggest that task-level suggestions provided important benefits but also introduced some challenges.  For example, task-level suggestions gave participants more new ideas that they needed to consider both individually and jointly, potentially increasing the level of effort required by the task.  Some of the challenges associated with the \textsc{BotTask} condition may have been alleviated had participants been given more time to complete each task.  While time constraints are common in interactive search studies, they can also influence perceptions, especially when participants \emph{do not adapt} their approach to a task based on the time available to complete it~\mbox{\cite{Crescenzi2021}}}.


\hl{Finally, during each task, participants had to gather information by interacting with the searchbot directly from the Slack channel.  In other words, participants could not search independently (e.g., using Google on their own browsers).  This decision was made to increase the level of engagement between the participants and the searchbot and analyze the effects of the searchbot condition on different types of outcomes.  Of course, in some situations, collaborators may be able to search on their own.  The impact of this decision is an open question for future work.  Future studies may want to extend our research to scenarios in which collaborators can both interact with an agent embedded in the communication channel as well as search on their own.}

\section{Conclusion}\label{sec:conclusion}

We reported on a Wizard of Oz study in which pairs of participants completed collaborative search tasks over the Slack messaging platform. To gather information, participants interacted with a conversational searchbot directly from the Slack channel. The role of the searchbot was played by a reference librarian. Participants in our study were exposed to three conditions, which varied based on the level of initiative the searchbot could take: (1) no initiative, (2) only dialog-level initiative (i.e., clarification questions), and (3) both dialog- and task-level initiative (i.e., clarification questions and task-level suggestions). Our six research questions examined the effects of the searchbot condition on different types of outcomes: (RQ1-RQ3) perceptions, (RQ4) behaviors, and (RQ5-RQ6) benefits and challenges reported by participants.

\hl{This research lies at the intersection of two areas of ongoing research: (1) collaborative search and (2) conversational search. Most research in conversational search has focused on designing systems to support a single searcher.  Less research has focused on designing conversational search systems to support \emph{multiple} collaborators working together on tasks that involve information seeking.  Additionally, most research has explored how systems can take dialog-level initiative by asking clarification questions to better address a user's information need.  In our study, we explored the potential benefits and pitfalls of a system that can take both dialog- \emph{and} task-level initiative to support collaborators.  Importantly, our results suggest that task-level suggestions (at times provided proactively) can provide important benefits but can also have negative consequences.  To name a few, proactive task-level suggestions can be disruptive, may require collaborators to expend \emph{unanticipated} effort, and can lead collaborators to perceive a loss of privacy and agency. Our results suggest that providing task-level suggestions is a challenging endeavor, even for a trained professional (i.e., a reference librarian) playing the role of the conversational agent. Based on our results, we have highlighted important directions for future work on mixed-initiative conversational search systems to support collaborations.}

\begin{acks}
This work was supported in part by Edward G. Holley research grant and NSF grant IIS-1451668. Any opinions, findings, conclusions, and recommendations expressed in this paper are the authors’ and do not necessarily reflect those of the sponsors. We also thank our anonymous reviewers for their feedback, and Danielle du Preez, Zhaopeng Xing, Austin Ward, and our reference librarians at UNC Chapel Hill for helping us with this study.
\end{acks}

\bibliographystyle{ACM-Reference-Format}
\bibliography{system_initiative_cscw}

\received{July 2021}
\received[revised]{November 2021}
\received[accepted]{November 2021}

\end{document}